\documentclass[12pt]{article}

\usepackage[a4paper,text={16.8cm,22.4cm}]{geometry}
\usepackage{amsmath,amsfonts,slashed,amssymb,tikz,bm,psfrag,graphicx,feynarts,color,dsfont}
\usepackage{multicol}
\usetikzlibrary{decorations.pathmorphing}
\usetikzlibrary{decorations.markings}
\usetikzlibrary{arrows,shapes}
\RequirePackage[sort&compress,square,comma,numbers]{natbib}
\allowdisplaybreaks 
\newcommand{\ord}{{\cal O}}
\newcommand{\e}{\epsilon}
\newcommand{\tx}{\text}
\newcommand{\da}{\dagger}
\newcommand{\La}{\mathcal{L}}
\newcommand{\bs}{\boldsymbol}
\newcommand{\tr}{\tx{Tr}} 
\newcommand{\UV}{\tx{UV}}
\newcommand{\TeV}{\tx{TeV}}
\newcommand{\SM}{\tx{SM}}
\newcommand{\RS}{\tx{RS}}
\newcommand{\NP}{\tx{NP}}
\newcommand{\Xb}{\boldsymbol{X}} 
\newcommand{\al}{\alpha}
\newcommand{\ga}{\gamma}
\newcommand{\Yb}{{\bs Y}}
\newcommand{\Q}{{\mathcal{Q}}}
\newcommand{\mkk}{M_\mathrm{KK}}
\newcommand{\Y}{\boldsymbol{Y}}

\tikzset{
vector/.style={decorate, decoration={snake,amplitude=2pt}, draw,segment length=6pt},
fermion/.style={draw=black, postaction={decorate}, decoration={markings,mark=at position .55 with {\arrow[draw=black]{>}}}},
fermionbar/.style={draw=black, postaction={decorate}, decoration={markings,mark=at position .55 with {\arrow[draw=black]{<}}}},
fermionnoarrow/.style={draw=black},
scalararrow/.style={dashed,draw=black, postaction={decorate}, decoration={markings,mark=at position .55 with {\arrow[draw=black]{>}}}},
scalar/.style={dashed}, 
gluon/.style={decorate, draw=black, decoration={coil,amplitude=4pt, mirror,segment length=7.85pt,aspect=0.8}},
vertex/.style={draw,circle,fill=black,inner sep=0pt,minimum size=1.2mm},  
comp/.style={line width=5pt, draw=blue!20},
elem/.style={line width=1pt, draw=black},
comp2/.style={line width=5pt, draw=blue!20},
elem2/.style={line width=1pt, draw=black},
ghost/.style={dotted,draw=black}
}

\begin{document}

\begin{titlepage}

\begin{flushright}
\normalsize
MITP/14-049\\
August 19, 2014
% v1: arXiv:1408.mmmm
\end{flushright}

\vspace{0.3cm}
\begin{center}
\Large\bf
Higgs Couplings and Phenomenology in a\\ 
Warped Extra Dimension 
\end{center}

\vspace{0.8cm}
\begin{center}
Raoul Malm$^a$, Matthias Neubert$^{a,b}$ and Christoph Schmell$^a$\\
\vspace{0.7cm} 
{\sl 
${}^a$PRISMA Cluster of Excellence \& Mainz Institute for Theoretical Physics\\
Johannes Gutenberg University, 55099 Mainz, Germany\\[3mm]
${}^b$Department of Physics, LEPP, Cornell University, Ithaca, NY 14853, U.S.A.}
\end{center}

\vspace{1.5cm}
\begin{abstract}
We present a comprehensive description of the Higgs-boson couplings to Standard Model fermions and bosons in Randall-Sundrum (RS) models with a Higgs sector localized on or near the infra-red brane. The analytic results for all relevant Higgs couplings including the loop-induced couplings to gluons and photons are summarized for both the minimal and the custodial RS model. The RS predictions for all relevant Higgs decays are compared with current LHC data, which already exclude significant portions of the parameter space. We show that the latest measurements are sensitive to KK gluon masses up to $20\,{\rm TeV}\times (y_\star/3)$ at 95\% confidence level for anarchic 5D Yukawa couplings bounded from above by $|(\bm{Y}_f)_{ij}|\le y_\star$. We also derive the sensitivity levels attainable in the high-luminosity run of the LHC and at a future linear collider. 
\end{abstract}
\vfil

\end{titlepage}

\section{Introduction}
\label{sec:intro}

In July 2012 the Higgs boson, the last missing piece of the Standard Model (SM), was discovered at the Large Hadron Collider (LHC) at CERN \cite{ATLAS:2012gk,CMS:2012gu}. Since then the hierarchy problem, i.e.\ the question about the mechanism that stabilizes the Higgs mass near the electroweak scale, is no longer a hypothetical issue. A promising possibility to solve the hierarchy problem is offered by Randall-Sundrum (RS) models \cite{Randall:1999ee}, in which the SM is embedded in a slice of anti-de Sitter space while the Higgs sector is localized on the ``infra-red (IR) brane'', one of two sub-manifolds bounding the extra dimension. The smallness of the electroweak scale can then be explained by the fundamental UV cutoff given by the warped Planck scale, whose value near the IR brane lies in the TeV range. Moreover, by allowing the fermion fields to propagate in the bulk, these models provide a natural explanation for the hierarchies observed in the flavor sector \cite{Grossman:1999ra,Gherghetta:2000qt,Huber:2000ie} and the smallness of flavor-changing neutral currents \cite{Agashe:2004ay,Agashe:2004cp,Csaki:2008zd,Casagrande:2008hr,Blanke:2008zb,Blanke:2008yr,Bauer:2009cf}.

The direct detection of Kaluza-Klein (KK) modes, massive copies of the SM particles with approximately equidistant mass gaps, would be a clear indication for a warped extra dimension. Unfortunately, none of these predicted particles have been observed yet, and electroweak precision measurements indicate that their masses could be too large for direct detection at the LHC. Thus, indirect searches like precision measurements of the Higgs-boson couplings to SM particles, which are accessible via studies of both the Higgs production cross sections and its various decay rates,  become an attractive alternative. In the context of Higgs physics, new-physics deviations from the SM can be searched for by measuring the signal rates 
\begin{equation}\label{eqn:RXXintro}
   R_X\equiv \frac{(\sigma\cdot{\rm BR)}(pp\to h\to X)_\NP}%
                  {(\sigma\cdot{\rm BR)}(pp\to h\to X)_\SM}
   = \frac{\sigma(pp\to h)_{\rm NP}}{\sigma(pp\to h)_{\rm SM}}\, 
    \frac{\Gamma(h\to X)_{\rm NP}}{\Gamma(h\to X)_{\rm SM}}\, 
    \frac{\Gamma_{h}^{\rm SM}}{\Gamma_{h}^{\rm NP}}
\end{equation}
for the production of the Higgs boson in $pp$ collisions at the LHC and its subsequent inclusive decay into an arbitrary final state $X$. Our work includes a detailed discussion of the signal rates $R_X$ for the most relevant decays into $X=b\bar b,\,\tau^+\tau^-,\,WW^*,\,ZZ^*$, and $\ga\ga$ in different incarnations of RS models. From~\eqref{eqn:RXXintro} we can read off that new physics can show up in three different ways. Firstly, it can lead to deviations in the Higgs production cross section $\sigma(pp\to h)$, which can be decomposed into the cross sections for Higgs production via gluon fusion, vector-boson fusion, Higgs-strahlung, and the associated production with a $t\bar t$ pair. The relative contributions read (for $m_h=125$\,GeV) \cite{Heinemeyer:2013tqa}
\begin{equation}
   \sigma(pp\to h) = 0.872\,\sigma_{ggh} + 0.070\,\sigma_{VVh} + 0.033\,\sigma_{Wh}
    + 0.020\,\sigma_{Zh} + 0.005\,\sigma_{t\bar th} \,.
\end{equation}
Secondly, new-physics effects can change the Higgs decay rates $\Gamma(h\to X)$, and thirdly  they can modify the total Higgs width $\Gamma_h$. Via the latter quantity the rates are sensitive to non-standard or invisible Higgs decays. In our analysis we take into account all three possibilities. While the gluon-fusion process has been discussed extensively in the literature \cite{Djouadi:2007fm,Falkowski:2007hz,Cacciapaglia:2009ky,Bhattacharyya:2009nb,Bouchart:2009vq,Casagrande:2010si,Azatov:2010pf,Goertz:2011hj,Carena:2012fk,Malm:2013jia}, we analyze the effects of the exchange of virtual KK resonances in the Higgs-strahlung and vector-boson fusion production processes for the first time. Moreover, we take a closer look at the Higgs decays into pairs of $W$ and $Z$ bosons, including their subsequent decays into leptons. This allows for a thorough discussion of the implications of the latest LHC results on the RS parameter space. 

In the context of various RS models, we summarize and discuss results for the various couplings of the Higgs boson to fermions and gauge bosons as well as the Higgs self-couplings. It has been reported in \cite{Peskin:2012we} that the LHC at $\sqrt{s}=14$\,TeV and with an integrated luminosity of 300\,fb$^{-1}$ has the potential to probe, in a model-independent way, deviations of the Higgs couplings to fermions in the range of $\sim 30\%$ and to gauge bosons in the range of $\sim 16\%$, both at 95\% confidence level (CL). At future lepton colliders like the International Linear Collider (ILC) \cite{Baer:2013cma,Klute:2013cx,Asner:2013psa,Tian:2013yda}, the sensitivity to deviations can be improved by almost one order of magnitude (assuming $\sqrt{s}=1$\,TeV and an integrated luminosity of 1000\,fb$^{-1}$). In order to explore to which extent it is possible to obtain evidence for models with warped extra dimensions by indirect measurements, we illustrate which regions of parameter space could be probed at these facilities.

We focus on RS models where the electroweak symmetry-breaking sector is localized on or near the IR brane. The extra dimension is chosen to be an $S^1/Z_2$ orbifold parametrized by a coordinate $\phi \in [-\pi, \pi]$, with two 3-branes localized on the orbifold fixed-points $\phi=0$ (UV brane) and $|\phi|=\pi$ (IR brane). The RS metric reads \cite{Randall:1999ee}
\begin{equation}
\label{eqn:RSmetric}
   ds^2 = e^{-2\sigma(\phi)}\,\eta_{\mu\nu}\,dx^\mu dx^\nu - r^2 d\phi^2
   = \frac{\epsilon^2}{t^2} \left( \eta_{\mu\nu}\,dx^\mu dx^\nu
    - \frac{1}{M_{\rm KK}^2}\,dt^2 \right) ,
\end{equation}
where $e^{-\sigma(\phi)}$, with $\sigma(\phi)=kr|\phi|$, is referred to as the warp factor. The size $r$ and curvature $k$ of the extra dimension are assumed to be of Planck size, $k\sim 1/r\sim M_{\rm Pl}$. The quantity $L=\sigma(\pi)=kr\pi$ measures the size of the extra dimension and is chosen to be $L\approx 33-34$ in order to explain the hierarchy between the Planck scale $M_{\rm Pl}$ and the TeV scale. We define the KK scale $M_{\rm KK}=k\epsilon$, with $\epsilon=e^{-\sigma(\pi)}$, which sets the mass scale for the low-lying KK excitations of the SM particles. On the right-hand side of \eqref{eqn:RSmetric} we have introduced a new coordinate $t=\epsilon\,e^{\sigma(\phi)}$, whose values on the UV and IR branes are $\epsilon$ and 1, respectively.\footnote{The dimensionless variable $t$ is related to the conformal coordinate $z$ frequently used in the literature by the simple rescaling $z=t/M_{\rm KK}\equiv R'\,t$.}  
In our analysis, we consider both the minimal and the custodially protected RS model, adopting the conventions and notations of \cite{Casagrande:2008hr,Casagrande:2010si}. In the minimal RS model the gauge group is taken to be $SU(3)_c\times SU(2)_L\times U(1)_Y$ like in the SM, and it is broken to $SU(3)_c\times U(1)_{\rm em}$ by the Higgs vacuum expectation model (vev). The RS model with custodial symmetry is based on the gauge group $SU(3)_c\times SU(2)_L\times SU(2)_R\times U(1)_X\times P_{LR}$, which is broken both on the UV and IR branes as described in detail in \cite{Agashe:2003zs,Csaki:2003zu,Agashe:2006at}. The discrete $P_{LR}$ symmetry sets the gauge couplings of the two $SU(2)$ symmetries equal to each other. We shall distinguish two scenarios of RS models, which differ in the localization of the Higgs sector. In models with a \textit{brane-localized Higgs field}, the inverse characteristic width of the Higgs field along the extra dimension $\Delta_h$ is assumed to be much larger than the inherent UV cutoff near the IR brane, i.e.~$\Delta_h\gg\Lambda_\TeV\sim\mbox{several}\,\mkk$ \cite{Hahn:2013nza}. In models in which the Higgs field lives in the bulk, the inverse width lies below the cutoff scale, and hence the structure of the Higgs profile can be resolved by the high-momentum modes of the theory. While the general bulk-Higgs case will be discussed in future work \cite{Archer:2014jca,BulkHiggs}, we only discuss the special case of models featuring a \textit{narrow bulk-Higgs field}, whose inverse width is such that $\mkk\ll\Delta_h\ll\Lambda_\TeV$ \cite{Malm:2013jia}. 

Our paper is structured as follows: In Section~\ref{sec:treeH} we calculate the cross sections for Higgs production via Higgs-strahlung and vector-boson fusion, as well as the decay rates of the Higgs boson into pairs of electroweak gauge bosons. In Section~\ref{sec:coupsA} we give a summary of the main Higgs couplings to fermions and gauge bosons in RS models, including the loop-induced couplings to two gluons and photons, and present expressions that are exact at first order in $v^2/\mkk^2$. A numerical study of both the CP-even and CP-odd Higgs couplings in the custodial RS model is performed in Section~\ref{sec:coupsN}. We comment on the possibility to detect deviations from the SM values of the Higgs couplings at the LHC operating at $\sqrt{s}=14$\,TeV and with an integrated luminosity of 300\,fb$^{-1}$, and an ILC operating at $\sqrt{s}=1$\,TeV with an integrated luminosity of 1000\,fb$^{-1}$. In Section~\ref{sec:bounds} we then compare the predictions for $pp\to h\to b\bar b$, $\tau^+\tau^-$, $WW^*$, $ZZ^*$, $\gamma\gamma$ obtained in the custodial RS scenario with the current data from the LHC, which can be used to deduce bounds on the relevant model parameters. Our main results are summarized in the conclusions.

\section{Higgs production and decay via $\bm{W}$ and $\bm{Z}$ bosons}
\label{sec:treeH}

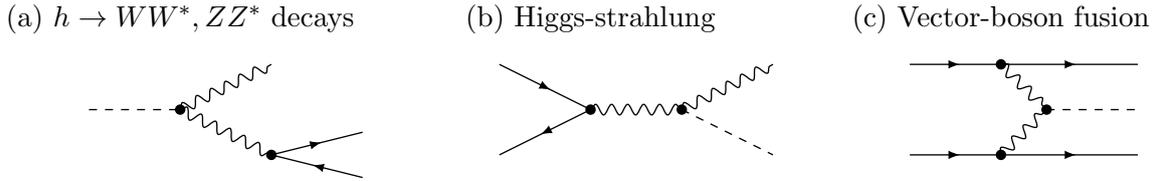
\begin{figure}[t!]
\begin{center}
\begin{tikzpicture}[line width=0.5pt,>=latex,scale=0.6]
\node (origin) at (0,0) {};
\begin{scope}[xshift=-9cm, yshift=0]
\draw[scalar] (-3,0) -- (-1,0);
\draw[vector] (-1,0) -- (1,1);
\draw[vector] (-1,0) -- (1,-1);
\draw[fermion] (1,-1) -- (3,-0.5);
\draw[fermion] (3,-1.5) -- (1,-1);
\draw (-1,0) node  [vertex] {};
\draw (1,-1) node  [vertex] {};
\draw (-1,2) node {\small (a) $h\to WW^*,ZZ^*$ decays};
\end{scope}
\begin{scope}[xshift=0cm, yshift=0]
\draw[vector] (-1,0) -- (1,0);
\draw[fermionbar] (-1,0) -- (-3,1);
\draw[fermion] (-1,0) -- (-3,-1);
\draw[vector] (1,0) -- (3,1);
\draw[scalar] (1,0) -- (3,-1);
\draw (-1,2) node {\small (b) Higgs-strahlung};
\draw (-1,0) node [vertex] {};
\draw (1,0) node [vertex] {};
\end{scope}
\begin{scope}[xshift=9cm, yshift=0]
\draw[fermion] (-3,1) -- (-1,1);
\draw[fermion] (-1,1) -- (2,1);
\draw[fermion] (-3,-1) -- (-1,-1);
\draw[fermion] (-1,-1) -- (2,-1);

\draw[vector] (-1,1) -- (0,0);
\draw[vector] (-1,-1) -- (0,0);
\draw[scalar] (0,0) -- (2,0);

\draw (-1,1) node [vertex] {};
\draw (-1,-1) node [vertex] {};
\draw (0,0) node [vertex] {};
\draw (-1,2) node {\small (c) Vector-boson fusion};
\end{scope}
\end{tikzpicture}
\parbox{15.5cm}
{\caption{\label{fig:treeHiggs}
Tree-level Feynman diagrams for the off-shell Higgs decays to pairs of $W$ and $Z$~bosons, and Higgs production in the Higgs-strahlung and vector-boson fusion processes.}}
\end{center}
\end{figure}

In this section we discuss in detail the structure of new-physics effects in the couplings of the Higgs boson to a pair of electroweak gauge bosons. These couplings are probed in the off-shell Higgs decays $h\to WW^*$ and $h\to ZZ^*$ with subsequent decays into four fermions, as well as in the production of the Higgs boson in vector-boson fusion or in the Higgs-strahlung process, see Figure~\ref{fig:treeHiggs}. These tree-level processes have in common that they involve the exchange of virtual vector bosons, which implies that in addition to the SM $W$ and $Z$ bosons we must consider the effect of the infinite towers of KK resonances. It is often assumed in the literature that the main effect of new physics on these processes arises from a rescaling of the on-shell $hVV$ couplings. We show that there are also several other effects that need to be accounted for, namely a possible rescaling of the Higgs vev, a modification of the couplings of the $W$ and $Z$ bosons to light fermions, and the exchange of new heavy particles in the off-shell propagators. In RS models all of these effects are indeed present, and accounting for them correctly will be important for a general definition of the signal strength in terms of the Higgs couplings to fermions and vector bosons in Section~\ref{sec:bounds}. To good approximation, however, we will show that the main effects can be accounted for by a multiplicative rescaling of the SM decay rates and production cross sections. For simplicity of presentation, the derivations in this section will be performed for the case of the minimal RS model. The extension to the case of the custodial model is presented in the Appendix.

\subsection{Higgs decay into vector bosons} 
\label{sec:hVV}

We begin by studying the decay of the Higgs boson to a pair of electroweak gauge bosons, taking $h\to WW^*$ as a concrete example. Since $m_h<2m_W$, this decay is only allowed if at least one of the $W$ bosons is produced off-shell. We thus consider the process $h\to W^- W^{+*}\to W^- f_i\bar f_j'$, where the off-shell boson decays into a pair of light fermions $f_i$ and $\bar f_j'$ with generation indices $i,j$. In the SM, the corresponding differential decay rate is given by \cite{Keung:1984hn}
\begin{equation}\label{dGds}
   \frac{d\Gamma}{ds} 
   = \frac{1}{16\pi^2 m_h^3}\,\frac{\Gamma(W^+\!\to\!f_i\bar f_j')}{m_W}\,\frac{m_W^2}{v^2}\,
    \frac{\lambda^{1/2}(m_h^2,m_W^2,s)}{\left( m_W^2-s \right)^2}
    \left[ \big( m_h^2-m_W^2 \big)^2 + 2s (5m_W^2-m_h^2) + s^2 \right] ,
\end{equation}
where $s$ is the invariant mass squared of the fermion pair, and $\lambda(x,y,z)=(x-y-z)^2-4yz$. We have expressed the result in terms of the on-shell decay rate for the process $W^+\!\to\!f_i\bar f_j'$,
\begin{equation}\label{GWff}
   \Gamma(W^+\to f_i\bar f_j') = N_c^f m_W\,\frac{g^2}{24\pi} \left| g_{ij,L} \right|^2 ,
\end{equation}
where $g$ denotes the $SU(2)_L$ gauge coupling, the color factor $N_c^f=1$ for leptons and~3 for quarks, and $g_{ij,L}=\delta_{ij}/\sqrt2$ for leptons and $V_{ij}^{\rm CKM}/\sqrt2$ for quarks. Performing the remaining integration over $s$ in the interval $0\le s\le(m_h-m_W)^2$ and neglecting fermion-mass effects, one obtains
\begin{equation}\label{GhZff}
   \Gamma(h\to W^- W^{+*}\to W^- f_i\bar f_j') 
   = \frac{m_h^3}{32\pi v^2}\,\frac{\Gamma(W^+\!\to\!f_i\bar f_j')}{\pi m_W}\,
    g\bigg( \frac{m_W^2}{m_h^2} \bigg) \,,
\end{equation}
where the first factor is one half of the (would-be) on-shell $h\to WW$ width in the limit $m_h\gg m_W$, the second factor accounts for the suppression due to the fact that one of the $W$ bosons in the decay $h\to WW^*$ is produced off-shell, and the phase-space function is given by
\begin{equation}
   g(x) = \frac{6x(1-8x+20x^2)}{\sqrt{4x-1}} \arccos\bigg(\frac{3x-1}{2x^{3/2}}\bigg)
    - 3x (1-6x+4x^2) \ln x - (1-x)(2-13x+47x^2) \,.
\end{equation}
The off-shell decay considered here arises if $x>1/4$. In the literature, it is common practice to define the off-shell $h\to WW^*$ decay rate as
\begin{equation}\label{hWW*}
   \Gamma(h\to W W^*)\equiv 2\sum_{f_i,f_j'}\,\Gamma(h\to W^+ f_i\bar f_j') \,,
\end{equation}
where the sum includes all fermion pairs with total mass lighter than $m_W$. The factor~2 accounts for the charge-conjugated decays $h\to W^-\bar f_i f_j'$. In the SM the expression for $\Gamma(h\to W W^*)$ has the same form as in (\ref{GhZff}), but with the partial decay rate $\Gamma(W^+\!\to\!f_i\bar f_j')$ replaced by twice the total decay width $\Gamma_W$ of the $W$ boson. 

Analogous formulas hold for the decays based on $h\to ZZ^*$, where we must replace $W\to Z$ everywhere and use the corresponding expression
\begin{equation}\label{GZff}
   \Gamma(Z\to f\bar f) 
   = N_c^f m_Z\,\frac{g^2}{24\pi c_w^2} 
    \left( g_{f,L}^2 + g_{f,R}^2 \right) ,
\end{equation} 
for the partial decay rates of the $Z$ boson in the SM, where $g_{f,L}=T_3^f-s_w^2\,Q_f$ and $g_{f,R}=-s_w^2\,Q_f$ are the left-handed and right-handed couplings of the various fermion species, and $s_w=\sin\theta_w$ and $c_w=\cos\theta_w$ are the sine and cosine of the weak mixing angle. In this case the total off-shell decay rate is defined as
\begin{equation}\label{hZZ*}
   \Gamma(h\to Z Z^*)\equiv \sum_f\,\Gamma(h\to Z f\bar f) \,,
\end{equation}
where the sum includes all fermions lighter than $m_Z/2$. It follows from this definition that for the golden channel
\begin{equation}
   \Gamma(h\to ZZ^*\to l^+ l^- l^+ l^-) 
   = \Gamma(h\to Z Z^*) \left[ \mbox{Br}(Z\to l^+ l^-) \right]^2 .
\end{equation}

We now discuss in detail how the above results must be modified in the context of the minimal RS model. For the purposes of this discussion it is convenient to define the weak mixing angle $s_w^2$ via the structure of the neutral current, which can be studied experimentally via the $Z$-pole polarization asymmetries observed at LEP. Alternative definitions are related to this one through the electroweak precision variables $S$, $T$ and $U$; see e.g.\ \cite{Archer:2014jca} for a detailed discussion. In the context of RS models one has $s_w^2=g_5^{\prime 2}/(g_5^2+g_5^{\prime 2})$ in terms of the 5D gauge couplings. If this ratio is extracted from experiment there are no new-physics corrections to the branching ratios $\mbox{Br}(W\to f_i\bar f_j')$ and $\mbox{Br}(Z\to f\bar f)$. Modifications arise for the Higgs couplings to vector bosons, the electroweak gauge couplings entering the partial decay rates (\ref{GWff}) and (\ref{GZff}), and due to the contributions of heavy KK resonances, which change the momentum-dependent gauge-boson propagator. Let us for concreteness consider the decay $h\to W^- W^{+*}$ to study the impact of these corrections in the context of the minimal RS model. In the Feynman diagram in Figure~\ref{fig:treeHiggs}(a) the off-shell gauge-boson propagator now contains the SM gauge boson and its infinite tower of KK excitations. The Feynman rule for the $W_\mu^{+(0)} W_\nu^{-(n)} h$ vertex is (with $n=0$ for the zero mode and $n>0$ for the KK excitations)
\begin{equation}\label{eqn:FRhWW}
   \frac{2i\tilde m_W^2}{v}\,\eta_{\mu\nu}\,2\pi\,\chi_0^W(1)\,\chi_n^W(1) \,,
\end{equation}
where $v$ denotes the Higgs vev in the RS model, which differs from the Higgs vev $v_{\rm SM}\equiv\big(\sqrt2 G_F\big)^{-1/2}$ by terms of order $v^2/\mkk^2$ \cite{Bouchart:2009vq,Malm:2013jia}. The quantity $\tilde m_W^2=\frac{g_5^2}{2\pi r}\,\frac{v^2}{4}$ is the leading contribution to the mass of the $W$ boson in an expansion in powers of $v^2/\mkk^2$, and $\chi_n^W(t)$ are the profiles of the $W$-boson KK modes along the extra dimension \cite{Casagrande:2008hr}. For the $W$-boson zero mode, one finds
\begin{equation}\label{eqn:chiW}
   \sqrt{2\pi}\,\chi_0^W(t) = 1 + \frac{m_W^2}{2\mkk^2} \left[ \frac12 - \frac{1}{2L} 
    - t^2 \left( L - \frac12 + \ln t \right) \right] + \dots \,,
\end{equation}
where here and below the ellipses denote terms of order $v^4/\mkk^4$ and higher. Note, in particular, that for the zero mode one encounters a correction factor \cite{Hahn:2013nza}
\begin{equation}\label{cWres}
   c_W = \frac{v_{\rm SM}}{v}\,\frac{\tilde m_W^2}{m_W^2}\,
    2\pi \left[ \chi_0^W(1) \right]^2
   = 1 - \frac{m_W^2}{2\mkk^2} \left( \frac{3L}{2} - 1 + \frac{1}{2L} \right) + \dots
\end{equation}
relative to the SM.

The Feynman rule for the $W_\mu^{+(n)}\bar u_A^{(i)} d_A^{(j)}$ vertex, where $A=L,R$ is a chirality label and $i,j$ labels the flavors of the SM quarks, is to an excellent approximation given by \cite{Casagrande:2008hr}
\begin{equation}
\label{eqn:Wqq}
   \frac{i}{\sqrt2}\,\frac{g_5}{\sqrt{2\pi r}}\,\sqrt{2\pi}\,\chi_n^W(\epsilon)\,
    V_{ij}^{\rm CKM}\,\gamma^\mu P_L \,,
\end{equation}
where $P_L=\frac12(1-\gamma_5)$ is a chiral projection operator. 
Corrections to this result, including the couplings to right-handed fermions, are strongly chirality suppressed. Note, in particular, that for the zero mode one encounters a correction factor
\begin{equation}\label{666}
   c_{\Gamma_W}^{1/2}\equiv \frac{g_5}{\sqrt{2\pi r} g}\,\sqrt{2\pi}\,\chi_0^W(\epsilon)
   = 1 - \frac{m_W^2}{2\mkk^2}\,\frac{1}{4L} + \dots
\end{equation}
relative to the SM, which will affect all decay amplitudes of the $W$ boson into light fermions.

It follows that, relative to the SM, we must make the following replacements in the SM decay amplitude for $h\to W^- W^{+*}\to W^- u_i\bar d_j$:
\begin{equation}\label{replace2}
   \frac{1}{m_W^2-s} \to
   \frac{v_{\rm SM}}{v}\,\frac{\tilde m_W^2}{m_W^2}\,\sqrt{2\pi}\,\chi_0^W(1)\,
    \frac{g_5}{\sqrt{2\pi r} g}\,2\pi\,B_W(1,\epsilon;-s) \,,
\end{equation}
where the quantity 
\begin{equation}
   2\pi\,B_W(t,t';-p^2)
   = \sum_{n\ge 0}\,\frac{2\pi\,\chi_n^W(t)\,\chi_n^W(t')}{\left(m_n^W\right)^2-p^2}
   = \frac{c_1(t,t')}{m_W^2-p^2} + \frac{c_2(t,t')}{2\mkk^2} + \dots 
\end{equation}
denotes the 5D gauge-boson propagator of the RS model, which has been calculated in closed form in \cite{Randall:2001gb,Csaki:2002gy,Hahn:2013nza}. In the last equation we show the first two terms in an expansion in powers of $v^2/\mkk^2$, valid under the assumption that $p^2<m_W^2$, which is appropriate for our analysis. The numerator structures are given by
\begin{equation} \label{eqn:c1c2}
\begin{aligned}
   c_1(t,t') &= 2\pi\,\chi_0^W(t)\,\chi_0^W(t') \,, \\
   c_2(t,t') &= L\,t_<^2 + \frac{1}{2L} + t^2 \left( \ln t - \frac12 \right)
    + t'^2 \left( \ln t'  - \frac12 \right) ,
\end{aligned}
\end{equation}
with $t_<=\min(t,t')$. At subleading order, we can now rewrite the right-hand side of (\ref{replace2}) in the form
\begin{equation}\label{replace}
   \frac{1}{m_W^2-s}\to c_{\Gamma_W}^{1/2}\,c_W \left[ \frac{1}{m_W^2-s}
    - \frac{1}{4\mkk^2} \left( 1 - \frac{1}{L} \right)
    + \dots \right] .
\end{equation}
This result has an intuitive form. The factor $c_{\Gamma_W}^{1/2}$ rescales the $W$-boson decay amplitudes of the SM in a uniform way, the factor $c_W$ rescales the Higgs-boson coupling to a $W^+ W^-$ pair, and the last term in brackets is the contribution of heavy KK resonances. Substituting the above expression for the gauge-boson propagator into (\ref{dGds}) and performing the integration over $s$, we obtain
\begin{equation}
\label{GhWWfin}
   \Gamma(h\to WW^*) = \frac{m_h^3}{16\pi v_{\rm SM}^2}\,
    \frac{c_{\Gamma_W} \Gamma_W^{\rm SM}}{\pi m_W}\,c_W^2 
    \left[ g\bigg( \frac{m_W^2}{m_h^2} \bigg) 
    - \frac{m_h^2}{2\mkk^2} \left( 1 - \frac{1}{L} \right) 
    h\bigg( \frac{m_W^2}{m_h^2} \bigg) + \dots \right] ,
\end{equation}
with
\begin{equation}
\begin{aligned}
   h(x) &= - (1-4x+12x^2)\,\sqrt{4x-1} \arccos\bigg(\frac{3x-1}{2x^{3/2}}\bigg) \\
   &\quad\mbox{}- \frac12 (1-6x+36x^2) \ln x + \frac16\,(1-x)(11-61x+38x^2) \,.
\end{aligned}
\end{equation}

The analysis of new-physics effects on the $h\to ZZ^*$ decay rate proceeds analogously. Instead of $c_W$ in (\ref{cWres}) one finds the correction factor
\begin{equation}\label{cZres}
   c_Z = \frac{v_{\rm SM}}{v}\,\frac{\tilde m_Z^2}{m_Z^2}\,
    2\pi \left[ \chi_0^Z(1) \right]^2
   = 1 - \frac{m_Z^2}{2\mkk^2} \left( L - 1 + \frac{1}{2L} \right) 
    - \frac{L m_W^2}{4\mkk^2} + \dots
\end{equation}
for the $hZZ$ coupling. Moreover, in the RS model the $Zf\bar f$ couplings entering the partial rates in (\ref{GZff}) get replaced by
\begin{equation}
\label{eqn:Zff}
   \frac{g}{c_w}\,g_{f,A}(s_w^2)
   \to \frac{g_5}{\sqrt{2\pi r}\,c_w}\,\sqrt{2\pi}\,\chi_0^Z(\epsilon)\,g_{f,A}(s_w^2) \,.
\end{equation} 
If the weak mixing angle is defined via the structure of the couplings $g_{f,A}(s_w^2)$, then the only difference with regard to the SM is a factor
\begin{equation}\label{777}
   c_{\Gamma_Z}^{1/2}\equiv \frac{g_5}{\sqrt{2\pi r} g}\,\sqrt{2\pi}\,\chi_0^Z(\epsilon)
   = c_{\Gamma_W}^{1/2} \left[ 1 + \frac{m_Z^2-m_W^2}{4\mkk^2}
    \left( 1 - \frac{1}{L} \right) + \dots \right] .
\end{equation}
Note that, if $m_Z$ and $s_w^2$ are taken as inputs, then the $W$-boson mass is a derived quantity, which obeys $m_W^2(m_Z,s_w^2)=m_Z^2 c_w^2 \left[ 1 + \frac{m_Z^2 s_w^2}{2\mkk^2} \left( L - 1 + \frac{1}{2L} \right) + \dots \right]$. As long as we choose $\mkk$ consistent with the bounds from electroweak precision tests (see Section~\ref{sec:coupsN}), this value will be consistent within errors with the measured $W$ mass.

The fact that the $L$-enhanced terms in the effective couplings $c_W$ in (\ref{cWres}) and $c_Z$ in (\ref{cZres}) are different is problematic from a phenomenological point of view, as this amounts to a breaking of custodial symmetry in the effective couplings of the Higgs to electroweak gauge bosons. Indeed, the difference $(c_W-c_Z)$ is related to the $T$ parameter, which receives dangerously large corrections in the minimal RS model \cite{Csaki:2002gy,Carena:2003fx}. Taming these effects has been the main motivation for the construction of RS models with a custodial symmetry in the bulk \cite{Agashe:2003zs,Csaki:2003zu,Agashe:2006at}. The extension of the above analysis to the RS scenario with a custodial symmetry is discussed in the Appendix. Here we shall briefly collect the relevant formulas for the various correction factors. The expressions for the correction factors to the $hVV$ vertices become 
\begin{equation}\label{eqn:cWZcust}
\begin{aligned}
   c_W \big|_{\rm cust} 
   &= 1 - \frac{m_W^2}{2\mkk^2} \left( 3L - 1 + \frac{1}{2L} \right) + \dots \,, \\
   c_Z \big|_{\rm cust} 
   &= 1 - \frac{m_W^2}{2\mkk^2} \left( 3L + 1 - \frac{1}{2L} \right) + \dots \,.
\end{aligned}
\end{equation}
Note that the custodial protection mechanism ensures that the leading, $L$-enhanced terms are now the same for both couplings \cite{Casagrande:2010si,Fichet:2013ola}, whereas the subleading terms are different. The correction factors $c_{\Gamma_{W,Z}}$ to the $W\to f\bar f'$ and $Z\to f\bar f$ decay rates remain unchanged. 

\subsection{Higgs-strahlung}
\label{sec:HV}

We now move on to study the cross section for the Higgs-strahlung process, in which the Higgs boson is produced in $pp$ collisions in association with a $W$ or $Z$ boson, see Figure~\ref{fig:treeHiggs}(b). Since the Feynman diagram for Higgs-strahlung is identical to that for the Higgs-boson decay into a pair of electroweak gauge bosons, it follows that the amplitude at the quark level receives exactly the same corrections as the Higgs decay amplitude discussed in the previous section. If we denote the invariant mass squared of the $hV$ pair in the final state by $s$, we immediately obtain from (\ref{replace}) (for $V=W,Z$) 
\begin{equation}
   \frac{d\sigma(pp\to hV)}{ds} = c_{\Gamma_V}\,c_V^2
    \left[ 1 + \frac{s-m_V^2}{2\mkk^2} \left( 1 - \frac{1}{L} \right) + \dots \right]
    \frac{d\sigma(pp\to hV)_{\rm SM}}{ds} \,.
\end{equation}
Because the $s$ dependence of the SM cross section is sensitive to the shapes of the parton distribution functions, it is not possible to derive a simple analytic formula for the corrections to the total Higgs-strahlung cross sections. However, the leading correction terms enhanced by $L$ are universal and independent of $s$. When only these terms are kept, one obtains 
\begin{equation}\label{eqn:cWh}
   \sigma(pp\to hV)\approx c_V^2\,\sigma(pp\to hV)_\SM \,.
\end{equation}
This approximation has been frequently used in the literature. In RS models it is accurate up to small corrections not enhanced by $L$.

\subsection{Higgs production in vector-boson fusion}
\label{sec:VBF}

We finally consider the vector-boson fusion process shown in Figure~\ref{fig:treeHiggs}(c). It involves two gauge-boson propagators, whose momenta we denote by $p_{1,2}$. In analogy with the discussion in the previous sections, we find that in order to account for new-physics effects one must replace
\begin{equation}
\begin{aligned}
\label{eqn:AWWh}
   \frac{1}{(m_V^2-p_1^2)\,(m_V^2-p_2^2)} 
   &\to \frac{v_{\rm SM}}{v}\,\frac{\tilde m_V^2}{m_V^2}\,
    \left( \frac{g_5}{\sqrt{2\pi r} g} \right)^2 
    (2\pi)^2\,B_V(1,\epsilon;-p_1^2)\,B_V(1,\epsilon;-p_2^2) \\
   &= \frac{c_{\Gamma_V}\,c_V}{(m_V^2-p_1^2)\,(m_V^2-p_2^2)} 
    \left[ 1 - \frac{2m_V^2-p_1^2-p_2^2}{4\mkk^2} \left( 1 - \frac{1}{L} \right) 
     + \dots \right]
\end{aligned}
\end{equation}
in the expression for the scattering amplitude. Once again the integrations over the virtual momenta flowing through the propagators cannot be performed in closed form, because they involve convolutions with parton distribution functions. However, the leading correction terms enhanced by $L$ are universal. When only these terms are kept, one obtains 
\begin{equation}
   \sigma(pp\to hqq')\approx c_V^2\,\sigma(pp\to hqq')_\SM \,,
\end{equation}
which is an approximation often adopted in the literature.

\section{Higgs couplings in RS models} 
\label{sec:coupsA}

In order to parameterize the RS contributions to the various Higgs couplings, we match them onto an effective Lagrangian defined at the electroweak scale $\mu\approx v$. For simplicity we neglect the effects of renormalization-group running from the new-physics scale $\mu\approx\mkk$ down to the electroweak scale, as their numerical impact is of minor importance. The phenomenologically most relevant Higgs couplings can be described using the following Lagrangian in the broken electroweak phase: 
\begin{equation}\label{eqn:effL}
\begin{aligned}
   \La_{\tx{eff}} &= c_W\,\frac{2 m_W^2}{v_\SM}\,h W_\mu^+ W^{-\mu} 
    + c_Z\,\frac{m_Z^2}{v_\SM}\,h Z_\mu Z^\mu 
    - \sum_{f=t,b,\tau} \frac{m_f}{v_\SM}\,h
    \bar f \left( c_f + c_{f5}\,i\gamma_5 \right) f \\[-3mm]
   &\quad\mbox{}- c_{3h}\,\frac{m_h^2}{2v_{\rm SM}}\,h^3 
    - c_{4h}\,\frac{m_h^2}{8v_{\rm SM}^2}\,h^4
    + c_g\,\frac{\al_s}{12\pi v_\SM}\,h G_{\mu\nu}^a G^{a,\mu\nu} 
    - c_{g5}\,\frac{\al_s}{8\pi v_\SM}\,h G_{\mu\nu}^a \Tilde G^{a,\mu\nu} \\
   &\quad\mbox{}+ c_\ga\,\frac{\al}{6\pi v_\SM}\,h F_{\mu\nu} F^{\mu\nu}
    - c_{\ga5}\,\frac{\al}{4\pi v_\SM}\,h F_{\mu\nu} \Tilde F^{\mu\nu} 
    + \dots \,. 
\end{aligned}
\end{equation}
We emphasize that it is not a complete list of operators. For instance, we have not included the operators $hZ_\mu\bar f\ga^\mu f$ and $hZ_\mu\bar f\ga^\mu\ga_5 f$ contributing to the $h\to ZZ^*\to Z\bar ff$ decay amplitude (and corresponding operators for $h\to WW^*$), since as shown in Section~\ref{sec:hVV} their contribution is subdominant. Furthermore, we do not consider the Higgs decay $h\to Z\ga$ or any flavor-violating couplings in this work. Both the CP-even couplings $c_i$ and the CP-odd coefficients $c_{i5}$ are real. In the SM $c_W=c_Z=c_f=c_{3h}=c_{4h}=1$ and $c_{f5}=c_g=c_{g5}=c_\ga=c_{\ga 5}=0$. 

\subsubsection*{Higgs couplings to fermions and electroweak gauge bosons}

In the SM, the Higgs boson couples to fermions and electroweak gauge bosons at tree level, with coupling strengths proportional to the masses of these particles. The non-universality of these couplings is the most distinguished feature of the Higgs mechanism. In RS models, modifications of the couplings arise from two effects: genuine corrections to the $hVV$ (with $V=W,Z$) and $h\bar f f$ vertices, and an overall rescaling of all couplings due to the shift of the Higgs vev, which appears because we use the SM vev $v_{\rm SM}$ in the effective Lagrangian (\ref{eqn:effL}). We now present explicit expressions for the various $c_i$ parameters, working consistently to first order in $v^2/\mkk^2$. Wherever possible, we will parameterize the differences between the minimal and the custodial RS model by means of a parameter $\xi$, which equals~1 in the minimal model and~2 in the custodial model. 

The Higgs couplings to $W$ and $Z$ bosons in RS models have been derived in \cite{Casagrande:2008hr,Casagrande:2010si,Hahn:2013nza} and given in {\eqref{cWres}, \eqref{cZres}, and (\ref{eqn:cWZcust})}. With $L\approx 33-34$, the $L$-enhanced contributions in these expressions are by far dominant numerically. Future precise measurements of $c_W$ and $c_Z$ would thus provide a direct tool to determine the ratio $\mkk/\sqrt{L}$ in the RS model.

The couplings of the Higgs boson to the third-generation fermions have been studied in detail in \cite{Casagrande:2010si}, where it was found that flavor-changing couplings are strongly suppressed. For the CP-even and CP-odd flavor-diagonal couplings, it follows that (with $f=t,b,\tau$ on the left-hand side and $f=u,d,e$ on the right-hand side)
\begin{equation}\label{eqn:hff}
   c_f +i c_{f5} 
   = 1 - \varepsilon_f - \frac{\xi L m_W^2}{4\mkk^2} 
    - \frac{\xi v^2}{3M_{\rm KK}^2}\,
    \frac{\big( \bm{Y}_f \bm{Y}_f^\dagger \bm{Y}_f \big)_{33}}{\big( \bm{Y}_f \big)_{33}}
    + \dots \,, 
\end{equation}
where $\bm{Y}_f$ denote the dimensionless, anarchic 5D Yukawa matrices in the up, down and lepton sectors. Note that the CP-odd couplings in (\ref{eqn:hff}) are solely due to the ``three-Yukawa terms''. The real-valued quantities $\varepsilon_f$ arise from overlap integrals of the ``wrong-chirality'' fermion profiles. They are given by 
\begin{equation}\label{eqn:epsf}
   \varepsilon_f = \left\{ \begin{array}{ll} 
    \big( \bs{\delta}_F \big)_{33} + \big( \bs{\delta}_f \big)_{33} \,; 
    & \mbox{minimal RS model,} \\[2mm]
    \big( \bs{\Phi}_F \big)_{33} + \big( \bs{\Phi}_f \big)_{33} \,;~
    & \mbox{custodial RS model.}
   \end{array} \right.
\end{equation}
Explicit expressions for the matrices $\bs{\delta}_{U,D,E}$ and $\bs{\delta}_{u,d,e}$ can be found in eq.~(5.13) of \cite{Casagrande:2008hr}, while those for the matrices $\bs{\Phi}_{U,D,E}$ and $\bs{\Phi}_{u,d,e}$ are given in eqs.~(6.19) and (6.20) of \cite{Casagrande:2010si}. They depend in a complicated way on the bulk mass parameters of the various 5D fermion fields. All of the quantities $\varepsilon_f$ are of ${\cal O}(v^2/\mkk^2)$, but in addition some of them are strongly chirality suppressed. For all practical purposes, one can retain $\varepsilon_u=(\bs{\delta}_U)_{33}+(\bs{\delta}_u)_{33}$ but approximate $\varepsilon_d\approx(\bs{\delta}_D)_{33}$, $\varepsilon_e\approx 0$, and similarly in the custodial model. Numerically, the $\varepsilon_f$ parameters turn out to play a numerically subleading role compared with the ``three-Yukawa terms'' in $c_f$. 

The Higgs couplings to the fermions do not only depend on the KK mass scale, but also on the dimensionless 5D Yukawa matrices. It is possible to simplify the Yukawa-dependent terms in the anarchic approach to flavor physics in RS models, in which the fundamental 5D Yukawa matrices are assumed to be structureless, and the observed hierarchies in the mass matrices of the SM fermions are explained in terms of their overlap integrals with the wave function of the Higgs scalar \cite{Grossman:1999ra,Gherghetta:2000qt,Huber:2000ie}. When scanning over the parameter space of an RS model, the various entries of the Yukawa matrices are taken to be complex random number subject to the condition that $|(\bs Y_f)_{ij}|\leq y_\star$, where the upper bound $y_\star={\cal O}(1)$ is a free parameter. For an ensemble of sufficiently many random matrices constructed in this manner, one can show that on average \cite{Malm:2013jia,Hahn:2013nza}
\begin{equation}\label{eqn:YYY}
   \bigg\langle \frac{\big( \bm{Y}_f\bm{Y}_f^\dagger\bm{Y}_f \big)_{33}}%
                     {\big( \bm{Y}_f \big)_{33}} \bigg\rangle
   = (2N_g-1)\,\frac{y_\star^2}{2} \,,
\end{equation}
where $N_g=3$ is the number of generations. It follows that the Higgs couplings to fermions are rather insensitive to the individual entries of the Yukawa matrices, but they do scale with $y_\star^2$. Hence, we encounter a similar situation as in the gauge-boson case, where the relevant parameter is now given by $\mkk/y_\star$. We should add at this point that in practice relation (\ref{eqn:YYY}) is subject to some flavor-dependent corrections, which arise when the scan over random Yukawa matrices is performed subject to the constraint that one obtains acceptable values for the quark and lepton masses and for the CKM matrix in the quark sector. When this is done, one finds numerically that the expectation value (\ref{eqn:YYY}) is slightly enhanced for the top quark and somewhat reduced for the bottom quark.\footnote{For $y_\star=1$, we find numerically that the expectation value (\ref{eqn:YYY}) is equal to 2.5 (as expected) for anarchic matrices, while it is 2.7 in the up-quark sector and 2.2 in the down-quark sector. We do not consider neutrino masses or the PMNS matrix in our analysis, since this would require the specification of the neutrino sector, which is both model dependent and of little relevance to Higgs physics.}

We close this subsection with a comment on a certain class of brane-Higgs models, in which one uses two different Yukawa matrices $\Y_f^C$ and $\Y_f^S$ in the Higgs couplings to the $Z_2$-even and $Z_2$-odd fermion fields. While in bulk-Higgs models the two matrices must be equal as a result of 5D Lorentz invariance, they can be different if the scalar sector is localized on the IR brane. We refer to models with $\Y_f^C\ne\Y_f^S$ as \textit{type-II brane-Higgs models}. In these scenarios, the Yukawa-dependent terms in \eqref{eqn:hff} change according to \cite{Malm:2013jia}
\begin{equation}\label{YCYSYC}
   \frac{\big( \bm{Y}_f \bm{Y}_f^{\dagger} \bm{Y}_f \big)_{33}}{\big( \bm{Y}_f \big)_{33}}
   \to \frac{\big( \bm{Y}_f^C \bm{Y}_f^{S\dagger} \bm{Y}_f^C \big)_{33}}%
            {\big( \bm{Y}_f^C \big)_{33}} \,.
\end{equation}
For the special case $\bm{Y}_f^S=0$, which was sometimes adopted in the literature, this term vanishes. There is then no contribution to the CP-odd couplings $c_{f5}$.

\subsubsection*{Higgs self-couplings}

One of the predictions of the SM is that the trilinear and quartic Higgs couplings can be expressed in terms of the Higgs-boson mass and the vev of the Higgs field, such that $c_{3h}=c_{4h}=1$ in (\ref{eqn:effL}). In RS models these coefficients receive calculable corrections, which for the minimal and the custodial RS models are described by the same formula in terms of the correction to the Higgs vev. As long as the Higgs sector is localized on or near the IR brane, one obtains \cite{Malm:2013jia}
\begin{equation}\label{Higgsself}
   c_{3h} = \frac{v_{\rm SM}}{v} = 1 - \frac{\xi L m_W^2}{4\mkk^2} + \dots \,, \qquad
   c_{4h} = \frac{v_{\rm SM}^2}{v^2} = 1 - \frac{\xi L m_W^2}{2\mkk^2} + \dots \,.
\end{equation}
For a KK mass scale of $\mkk=1.5$\,TeV, one finds a 2.4\% (4.8\%) reduction of the trilinear coupling and a 4.8\% (9.6\%) reduction of the quartic coupling in the minimal (custodial) RS model. We mention that moving the Higgs field into the bulk would attenuate these deviations and move the couplings closer to their SM values \cite{BulkHiggs}. Such small deviations will not be measurable by the LHC, and even for a future linear collider like the ILC this is probably out of reach. Therefore, we refrain from presenting a detailed numerical analysis of the Higgs self-couplings in the subsequent section. 

\subsubsection*{Loop-induced Higgs couplings to two gluons}

In the SM, the Higgs boson couples to massless gluons and photons only via loop diagrams containing heavy SM particles. Direct couplings, such as the ones contained in the effective Lagrangian (\ref{eqn:effL}), are absent in the SM. In the context of RS models such direct couplings are induced at one-loop order via the exchange of heavy KK resonances. We begin with a discussion of the loop-induced Higgs couplings to gluons, which are relevant for the calculation of the gluon-fusion cross section $\sigma(gg\to h)$, which is the main Higgs production channel at high-energy hadron colliders such as the LHC. In the present work we concentrate on the case of the Higgs sector being localized near the IR brane, which has been discussed in several works \cite{Djouadi:2007fm,Falkowski:2007hz,Cacciapaglia:2009ky,Bhattacharyya:2009nb,Bouchart:2009vq,Casagrande:2010si,Azatov:2010pf,Goertz:2011hj,Carena:2012fk,Malm:2013jia}. As mentioned in the Introduction, the result for the contribution of the infinite tower of KK resonances exhibits a UV sensitivity in the sense that it is sensitive to the precise nature of the localization mechanism. 

In the limit where we neglect ${\cal O}(v^2/\mkk^2)$ corrections which in addition are strongly chirality suppressed, the expressions for the induced Higgs couplings to two gluons read
\begin{equation}\label{eqn:cg}
   c_g + ic_{g5} = \left\{ \begin{array}{ll} 
    \tr\,g(\bs X_u) + \tr\,g(\bs X_d) + \varepsilon_u + \varepsilon_d \,; 
    & \mbox{minimal RS model,} \\[2mm]
    \tr\,g(\sqrt2\bs X_u) + 3\,\tr\,g(\sqrt2\bs X_d) + \varepsilon_u + \varepsilon_d \,;~
    & \mbox{custodial RS model.}
   \end{array} \right.
\end{equation}
The quantities
\begin{equation}\label{eqn:Xq}
   \bs X_f = \frac{v}{\sqrt2\mkk} \sqrt{\Yb_f \Yb_f^\da}
\end{equation} 
are entirely given by the dimensionless 5D Yukawa matrices of the RS model. Note that the Yukawa matrices are the same in both the minimal and the custodial RS model, but there is an additional $\sqrt2$ in the argument of the function $g(\bs X_f)$ in the latter case. For the two scenarios with a brane-localized and a narrow bulk-Higgs sector, one finds \cite{Carena:2012fk,Malm:2013jia}
\begin{equation}\label{eqn:gfunc}
\begin{aligned}
   g(\bm{X}_f) \big|_{\rm brane\;Higgs} 
   &= - \frac{\bm{X}_f\tanh\bm{X}_f}{\cosh2\bm{X}_f}
    = - \frac{v^2}{2\mkk^2}\,\bm{Y}_f\bm{Y}_f^\dagger
     + \dots \,, \\ 
   g(\bm{X}_f) \big|_{\rm narrow\;bulk\;Higgs} 
   &= \bm{X}_f\tanh\bm{X}_f = \frac{v^2}{2\mkk^2}\,\bm{Y}_f\bm{Y}_f^\dagger
    + \dots \,,
\end{aligned}
\end{equation}
so that the effect from the KK tower is approximately equal but of opposite sign in the two scenarios. For a large ensemble of random matrices, one obtains on average \cite{Malm:2013jia,Hahn:2013nza}
\begin{equation}\label{eqn:TrYY}
   \Big\langle \tr\,\bm{Y}_f\bm{Y}_f^\dagger \Big\rangle
   = N_g^2\,\frac{y_\star^2}{2} \,.
\end{equation}
Due to the additional factors $\sqrt2$ and~3 in the second case in (\ref{eqn:cg}), the quark KK tower contribution in the custodial RS model is roughly four times larger than in the minimal RS model. Note that with the hermitian matrices $\Xb_f$ the traces over the matrix-valued functions $g(\Xb_f)$ are real, so that
\begin{equation}
   c_{g5} = 0 \,,
\end{equation}
irrespective of the Higgs localization or the type of RS model (minimal or custodial). For the type-II brane-Higgs model, the function $g(\bm{X}_f)$ in the first line of (\ref{eqn:gfunc}) must be replaced by $-\frac{v^2}{2\mkk^2}\,\bm{Y}_f^C\bm{Y}_f^{C\dagger}+\dots$ \cite{Malm:2013jia}, and hence to leading order there is no difference with the result shown above. In this model the CP-odd coupling $c_{g5}$ receives contributions starting at ${\cal O}(v^4/\mkk^4)$, which are however too small to be of any phenomenological significance. In the subsequent sections we will therefore restrict ourselves to a study of the two cases shown in (\ref{eqn:gfunc}).

When the top-quark is integrated out from the effective Lagrangian (\ref{eqn:effL}), additional contributions to the effective $hgg$ couplings are induced at one-loop order. They can be accounted for by introducing the effective coefficients
\begin{equation}\label{eqn:cgeff}
   c_g^{\rm eff} = \frac{c_g + A_q(\tau_t)\,c_t}{A_q(\tau_t)}\,, \qquad
   c_{g5}^{\rm eff} = \frac{c_{g5} + B_q(\tau_t)\,c_{t5}}{A_q(\tau_t)} \,,
\end{equation}
which we have normalized such that $c_g^{\rm eff}=1$ in the SM. Explicit expressions for the top-quark loop functions $A_q(\tau_t)\approx 1.03$ and $B_q(\tau_t)\approx 1.05$ (with $\tau_t=4m_t^2/m_h^2$) can be found, e.g., in \cite{Beneke:2002jn,Djouadi:2005gj}. Both approach 1 for $\tau_t\to\infty$, and it is an excellent approximation to use the asymptotic values for the small new-physics corrections to the Wilson coefficients. It then follows that the terms proportional to $\varepsilon_u$, which in $c_g^{\rm eff}$ combine to $\varepsilon_u\big[1-A_q(\tau_t)\big]$, can be safely neglected. Note also that to a very good approximation $c_{g5}^{\rm eff} \approx c_{t5}$. 

\subsubsection*{Loop-induced Higgs couplings to two photons}

We finally turn our attention to the couplings of the Higgs boson to two photons, which play a crucial role for the $h\to\ga\ga$ decay channel, in which the Higgs boson has been discovered in 2012. Neglecting as before ${\cal O}(v^2/\mkk^2)$ corrections which in addition are strongly chirality suppressed, the expressions for the induced Higgs couplings to two photons in the minimal RS model read \cite{Hahn:2013nza}
\begin{equation}\label{cgacga5}
   c_\ga + ic_{\ga 5} = N_c Q_u^2\,\big[ \tr\,g(\bs X_u) + \varepsilon_u \big] 
    + N_c Q_d^2\,\big[ \tr\,g(\bs X_d) + \varepsilon_d \big]
    + Q_e^2\,\tr\,g(\bs X_e) - \frac{21}{4}\,\nu_W \,, 
\end{equation}
while in the custodial model one obtains
\begin{equation}\label{cgacga5c}
\begin{aligned}
   c_\ga + ic_{\ga 5} 
   &= N_c Q_u^2\,\tr\,g(\sqrt2\bs X_u) 
    + N_c\!\left( Q_u^2 + Q_d^2 + Q_\lambda^2 \right) \tr\,g(\sqrt2\bs X_d) 
    + Q_e^2\,\tr\,g(\bs X_e) \\
   &\quad \mbox{}+ N_c Q_u^2\,\varepsilon_u + N_c Q_d^2\,\varepsilon_d - \frac{21}{4}\,\nu_W \,.
\end{aligned}
\end{equation}
They receive KK contributions from the quark and lepton loops as well as from loops of $W$ bosons and scalar Goldstone fields. Here $Q_{u,d,e}$ denote the electric charges of the SM fermions, and $Q_\lambda=\frac53$ is the charge of a new exotic, heavy fermion species encountered in the custodial RS model. The precise embeddings of the SM quark fields into the extended gauge symmetry has been discussed in detail in \cite{Albrecht:2009xr,Casagrande:2010si}. For the lepton fields two types of embeddings have been studied in \cite{Hahn:2013nza}. Here we adopt the simplest assignment, in which the left-handed neutrino and electron are put into an $SU(2)_L$ doublet (as in the SM) and the right-handed electron along with a new, exotic neutral particle $N_R$ into an $SU(2)_R$ doublet. The infinite tower of the KK excitations of the $W$ bosons (including the Goldstone fields) contributes \cite{Bouchart:2009vq,Casagrande:2010si,Hahn:2013nza}
\begin{equation}\label{eqn:nuW}
   \nu_W = \frac{m_W^2}{2\mkk^2} \left( \xi L - 1 + \frac{1}{2L} \right) + \dots \,.
\end{equation}
Like in the case of the gluon-fusion channel $gg\to h$, we defined effective coefficients obtained after the heavy particles $t$, $W$ and $Z$ of the SM have been integrated out. They are related to the above coefficients by
\begin{equation}\label{eqn:cga5}
   c_\ga^{\rm eff} 
   = \frac{c_\ga + N_c Q_u^2\,A_q(\tau_t)\,c_t - \frac{21}{4}\,A_W(\tau_W)\,c_W}%
          {N_c Q_u^2\,A_q(\tau_t) - \frac{21}{4}\,A_W(\tau_W)} \,,
    \qquad
   c_{\ga 5}^{\rm eff} 
   = \frac{c_{\ga 5} + N_c Q_u^2\,B_q(\tau_t)\,c_{t5}}%
     {N_c Q_u^2\,A_q(\tau_t) - \frac{21}{4}\,A_W(\tau_W)} \,,
\end{equation}
where again we have chosen the normalization such that $c_\gamma^{\rm eff}=1$ in the SM. The explicit form of the $W$-boson loop function $A_W(\tau_W)\approx 1.19$ (with $\tau_W=4m_W^2/m_h^2$), which approaches~1 for $\tau_W\to\infty$, can be found in  \cite{Beneke:2002jn,Djouadi:2005gj}. From the fact that the coefficient $c_{\ga 5}$ in (\ref{cgacga5}) and (\ref{cgacga5c}) vanishes, it follows that to a very good approximation 
\begin{equation}\label{cga5rela}
   c_{\ga 5}^{\rm eff} \approx -0.28\,c_{t5} \,.
\end{equation}

\section{Numerical analysis of Higgs couplings}
\label{sec:coupsN}

We now study the structure of new-physics effects to both tree-level and loop-induced Higgs couplings to fermions and gauge bosons in the context of the RS model with custodial symmetry, for which the bounds derived from electroweak precision tests allow for KK masses in the few TeV range. For example, a recent tree-level analysis of the $S$ and $T$ parameters yields $M_{g^{(1)}}>4.8$\,TeV (at 95\% CL) for the mass of the lightest KK gluon and photon resonances \cite{Malm:2013jia}, and somewhat lighter masses are possible for the KK fermion resonances \cite{Carena:2006bn,Cacciapaglia:2006gp,Contino:2006qr}. We will see that these bounds still allow for sizable effects in the Higgs sector. On the other hand, the corresponding bound $M_{g^{(1)}}>12.3$\,TeV (at 95\% CL) obtained in the minimal RS model is so high that the resulting corrections to the Higgs couplings are generally below the sensitivity level of present and planned collider experiments. In our analysis we take $m_h=125.6$\,GeV for the Higgs mass and $m_t=172.6$\,GeV for the pole mass of the top quark. The parameter $L=\ln(M_{\rm Pl}/\Lambda_{\rm TeV})$ is chosen to be $L=33.5$. 

\subsubsection*{Tree-level Higgs couplings}

In the custodial RS model, the corrections to the tree-level Higgs couplings to $W$ and $Z$ bosons in (\ref{eqn:cWZcust}) are identical up to very small corrections not enhanced by $L$. Introducing the mass $M_{g^{(1)}}\approx 2.45\,\mkk$ of the lightest KK gluon instead of the KK scale $\mkk$, which is independent of the details of the localization of the scalar sector and the choice of the electroweak gauge group \cite{Davoudiasl:1999tf}, we obtain
\begin{equation}\label{cWcZres}
   c_W\approx c_Z\approx 1 - 0.078 \left( \frac{5\,\mbox{TeV}}{M_{g^{(1)}}} \right)^2 .
\end{equation}
Realistically, with KK masses not in conflict with electroweak precision tests, we might thus expect corrections of a few up to a maximum of 10\%. The corrections to the Higgs self-couplings in (\ref{Higgsself}) are even smaller; the coefficients in front of the correction term are 0.026 for $c_{3h}$ and 0.052 for $c_{4h}$.

\begin{figure}
\begin{center}
\psfrag{x}[]{\small $M_{g^{(1)}}~{\rm [TeV]}$}
\psfrag{y}[b]{\small $c_t$}
\psfrag{z}[b]{\small $c_{t5}$}
\psfrag{w}[b]{\small custodial RS model}
\psfrag{a}[b]{\hspace{1cm} \scriptsize $y_\star=0.5$}
\psfrag{b}[b]{\hspace{1.03cm} \scriptsize $y_\star=1.5$}
\psfrag{c}[b]{\hspace{0.79cm} \scriptsize$y_\star=3$}
\includegraphics[width=\textwidth]{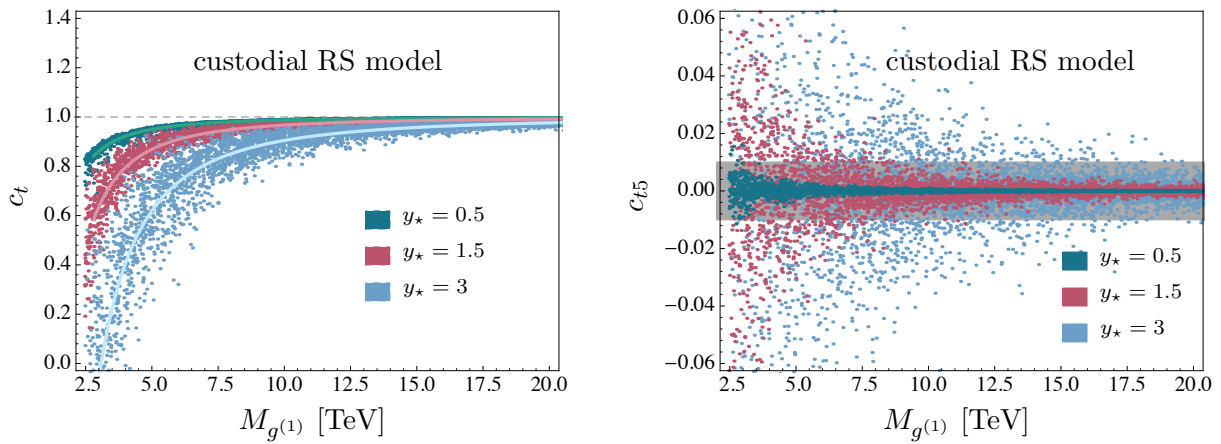} 
\parbox{15.5cm}
{\caption{\label{fig:ctct5}
Predictions for the Higgs couplings to top quarks as a function of the KK gluon mass $M_{g^{(1)}}$ in the custodial RS model. The green, red, and blue scatter points correspond to model points obtained using $y_\star=0.5$, 1.5, and 3, respectively. The overlaid lines in the left plot show fits to the various distributions as explained in the text. The gray band in the right plot shows the experimental bound on $|c_{t5}|$ derived from the electron EDM (at 90\% CL).}}
\end{center}
\end{figure}

Next we study the corrections to the CP-even and CP-odd Higgs couplings $c_f$ and $c_{f5}$ to the third-generation fermions, as obtained from (\ref{eqn:hff}). In analogy to our previous analyses in \cite{Malm:2013jia,Hahn:2013nza}, we generate three sets of 5000 random and anarchic 5D Yukawa matrices, whose entries satisfy $|(\bm{Y}_q)_{ij}| \leq y_\star$ with $y_\star=0.5$, 1.5, and 3, and which correctly reproduce the Wolfenstein parameters $\bar\rho$ and $\bar\eta$ of the unitarity triangle. Furthermore, we choose the bulk mass parameters $c_{Q_i}<1$ and $c_{q_i}<1$ such that we reproduce the correct values for the SM quark masses evaluated at the scale $\mu=1$\,TeV. Figure~\ref{fig:ctct5} shows the Higgs couplings to top quarks as a function of the mass of the lightest KK gluon state and for three different values of $y_\star$. In accordance with (\ref{eqn:hff}) and (\ref{eqn:YYY}) we observe that $c_t$ is reduced compared to the SM value~1 for almost all parameter points, where the depletion increases with larger values of $y_\star$. The corresponding plots for $c_b$ and $c_\tau$ would look very similar, with the magnitude of the corrections somewhat reduced. The main difference is due to the different values of the $\varepsilon_f$ parameters in the three cases, but their numerical impact is subleading. The solid lines in the left plot in the figure show simple polynomial fits of the form $c_f=1-a_f\,(5\,{\rm TeV}/M_{g^{(1)}})^2$ to the scatter points, with coefficients $a_f=a_f(y_\star)$ given in Table~\ref{tab:aibi}. We like to add a brief comment concerning the type-II brane Higgs model at this point, in which the three-Yukawa terms must be replaced according to (\ref{YCYSYC}) and have a vanishing expectation value. While the remaining terms in (\ref{eqn:hff}) still give rise to small negative corrections, the corresponding scatter plots would show points scattered more or less around the central value $c_i=1$, and which can become larger than 1 for not too small values for $y_\star$ due to the indefinite sign of the three-Yukawa terms. Although they are not as pronounced as in the conventional brane-Higgs scenarios, significant effects on the Higgs coupling to the top quark are still possible. For example, with $y_\star=3$ a modification of $c_t$ by 20\% is possible for KK excitations as heavy as 7.5\,TeV.

\begin{table}
\begin{center}
\begin{tabular}{|c|ccc|}
\hline
$y_\star$ & 0.5 & 1.5 & 3 \\ 
\hline
$a_t$ & 0.050 & 0.131 & 0.381 \\
$a_b$ & 0.033 & 0.085 & 0.243 \\ 
$a_\tau$ & 0.030 & 0.076 & 0.223 \\
\hline
\end{tabular}
\parbox{15.5cm}
{\caption{\label{tab:aibi}
Fit coefficients $a_f$ for different values of $y_\star$.}}
\end{center}
\end{table} 

The CP-odd couplings of the Higgs to two fermions $c_{f5}$ in the RS model are given by the second expression in \eqref{eqn:hff}. For random complex Yukawa matrices with entries bounded by $|(\bs Y_f)_{ij}|\le y_\star$, we find an approximately Gaussian distribution with zero mean and non-Gaussian tails, which can be reduced by imposing a lower bound on the magnitude of $\big|(\bm{Y}_f)_{33}\big|$. In the vicinity of the peak the distribution is approximately normal, with standard deviation
\begin{equation}
   \sigma_{c_{f5}} \approx \frac{v^2\,y_\star^2}{3\mkk^2}
   \approx 0.044 \left( \frac{y_\star}{3} \right)^2
    \left( \frac{5\,\mbox{TeV}}{M_{g^{(1)}}} \right)^2 .
\end{equation}
Due to the constraint that we must obtain realistic values of the quark masses and CKM mixing angles the actual results differ slightly from this result. It has been argued in \cite{Brod:2013cka} that present experimental bounds on electric dipole moments (EDMs) of the electron, neutron and mercury impose non-trivial bounds on the CP-odd Higgs couplings to the third-generation fermions. The strongest constraint exists for the magnitude on $c_{t5}$ and comes from the EDM of the electron, which is sensitive to the $h t\bar t$ couplings via two-loop Barr-Zee diagrams. Using the present 90\% CL upper limit $d_e<8.7\cdot 10^{-29} e\,\mbox{cm}$ \cite{Baron:2013eja} and assuming that the Higgs coupling to electrons is not changed with respect to its SM value, one obtains $|c_{t5}|<0.01$ \cite{Brod:2013cka}. In the RS models considered in this work this assumption is valid to high accuracy, since corrections to the $he^+ e^-$ coupling are strongly chirality suppressed. This resulting bound is shown by the gray band in the right plot in Figure~\ref{fig:ctct5}. Interestingly, we find that for $y_\star\gtrsim 1$ there are many points in RS parameters space for which $c_{t5}$ takes values of the same order of magnitude as the experimental bound. Hence, in the context of RS models it is conceivable that first hints of a non-zero electron EDM might be seen in the next round of experiments.

\subsubsection*{Loop-induced Higgs couplings}

\begin{figure}[t!]
\begin{center}
\psfrag{x}[]{\small $M_{g^{(1)}}~{\rm [TeV]}$}
\psfrag{y}[b]{\small $c_\ga^{\rm eff}$}
\psfrag{z}[]{\small \hspace{3mm} \begin{tabular}{l} custodial RS model \\[-1mm] narrow bulk Higgs \end{tabular}}
\psfrag{w}[]{\small \begin{tabular}{l} custodial RS model \\[-1mm] brane Higgs \end{tabular}}
\psfrag{a}[b]{\hspace{1cm} \scriptsize $y_\star=0.5$}
\psfrag{b}[b]{\hspace{1.03cm} \scriptsize $y_\star=1.5$}
\psfrag{c}[b]{\hspace{0.79cm} \scriptsize$y_\star=3$}
\includegraphics[width=0.96\textwidth]{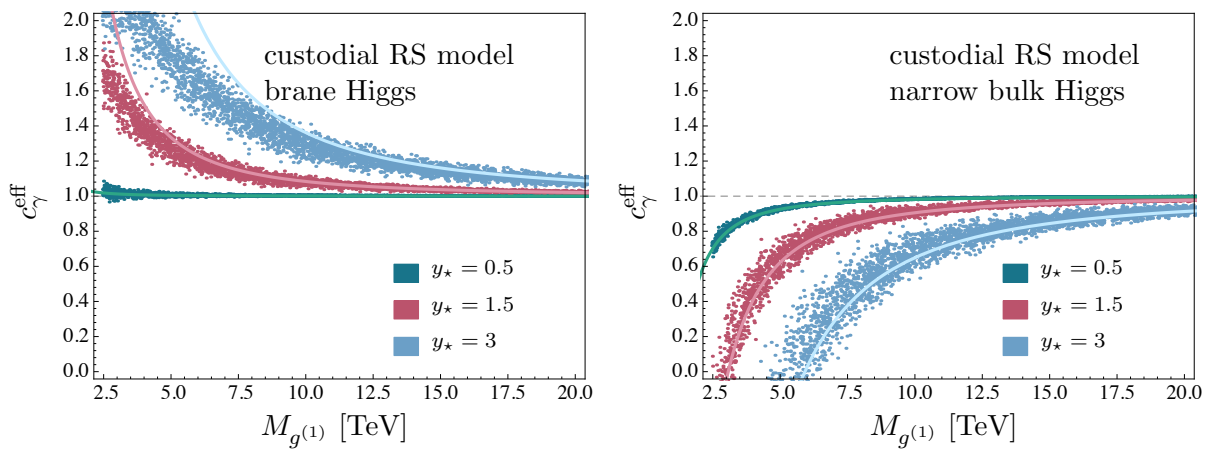} 
\parbox{15.5cm}
{\caption{\label{fig:cgammaeff}
Predictions for the CP-even effective Higgs coupling to two photons as a function of the KK gluon mass $M_{g^{(1)}}$ in the custodial RS model, for the scenarios with a brane-localized scalar sector (left) and a narrow bulk-Higgs field (right). The green, red, and blue scatter points correspond to model points obtained using $y_\star=0.5$, 1.5, and 3, respectively. The overlaid lines show the approximate results (\ref{cgeffcgaeff}).}}
\end{center}
\end{figure}

We now move on to study the loop-induced $hgg$ and $h\ga\ga$ couplings in the custodial RS model. They are of special interest, since they are very sensitive probes of the effects of virtual KK resonances. We concentrate on the CP-even couplings $c_g^{\rm eff}$ and $c_\ga^{\rm eff}$, since current measurements are not sufficiently precise to probe the CP-odd couplings.\footnote{There exist proposals for how to probe $c_{\gamma5}^{\rm eff}$ in $h\to\ga\ga$ decays in which both photons undergo nuclear conversion, by measuring certain kinematic distributions of the electron-positron pairs \cite{Bishara:2013vya}. Unfortunately, however, the level of sensitivity one can achieve does not allow one to probe the very small effects (\ref{cga5rela}) predicted in RS models, where the CP-odd $ht\bar t$ coupling is the only source of the effect.} 
Using the explicit expressions for $c_g^{\rm eff}$ and $c_\ga^{\rm eff}$ in \eqref{eqn:cgeff} and \eqref{eqn:cga5}, it is straightforward to derive approximate expressions for these coefficients which help to understand the interplay of the various contributions. To this end, we expand the fermion KK tower contributions in \eqref{eqn:cg} and \eqref{cgacga5c} to first order in $v^2/\mkk^2$ and employ (\ref{eqn:YYY}) and (\ref{eqn:TrYY}). We also approximate the top-quark loop function $A_q(\tau_t)$ by its asymptotic value~1 and neglect subleading terms not enhanced by $L$ in the bosonic contributions. This yields
\begin{equation}\label{cgeffcgaeff}
\begin{aligned}
   c_g^{\rm eff}  
   &\approx  1  + \frac{v^2}{2\mkk^2} \left[ 
    \left( \mp 36 - \frac{10}{3} \right) y_\star^2 - \frac{L m_W^2}{v^2} \right] 
   \approx 1 + \frac{v^2}{2\mkk^2} \left[ \left( \mp 36.0 - 3.3 \right) y_\star^2 - 3.6 \right] \\ 
   c_\ga^{\rm eff} &\approx 1 + \frac{v^2}{2\mkk^2} \left[
    \frac{1}{|C_\ga^\SM|} \left( \pm \frac{213}{2} + \frac{40}{9} \right) y_\star^2 
    - \frac{ 21(A_W(\tau_W) - 1)}{2|C_\ga^\SM|}\,\frac{L m_W^2}{v^2} 
    - \frac{L m_W^2}{v^2} \right] \\
   &\approx 1 + \frac{v^2}{2\mkk^2} \left[ \left( \pm 21.7 + 0.9 \right) y_\star^2 - 5.1\right] .
\end{aligned}
\end{equation}
Here the upper sign holds for the brane-Higgs case, while the lower one corresponds to the narrow bulk-Higgs scenario. We have kept the dependence on the one-loop SM amplitude $C_\ga^{\rm SM}= \frac{4}{3}-\frac{21}{4} A_W(\tau_W)\approx -4.9$ explicit. In each square bracket, the first term is due to the effects of KK fermion resonances, while the second term accounts for the vev shift and the contribution of bosonic KK states (for $c_\ga^{\rm eff}$). The fermionic contributions enter the two coefficients with opposite signs and are larger in magnitude in the case of $c_g^{\rm eff}$. Figure~\ref{fig:cgammaeff} shows our predictions for the coefficient $c_\ga^{\rm eff}$ as a function of the mass of the lightest KK gluon resonance and for different values of $y_\star$. We recall the well-known fact that the results exhibit a large sensitivity to the precise nature of the localization of the scalar sector on or near the IR brane. On average, the distributions of scatter points follow the approximate formulas shown in (\ref{cgeffcgaeff}); however, in the brane-Higgs case higher-order corrections become important for small $M_{g^{(1)}}$ values, and they are included in our phenomenological analysis below. The corresponding information on how $c^{\rm eff}_g$ depends on $M_{g^{(1)}}$ and $y_\star$ can be deduced from the correlation between the two loop-induced couplings, to which we turn now. 

\subsubsection*{Correlations between Higgs couplings}

\begin{figure}
\begin{center}
\psfrag{x}[]{\small $c_t$}
\psfrag{y}[b]{\small $c_b$}
\psfrag{z}[b]{\hspace{10mm} \small custodial RS model}
\psfrag{a}[b]{\hspace{1cm}  \scriptsize $y_\star=0.5$}
\psfrag{b}[b]{\hspace{1.03cm}  \scriptsize $y_\star=1.5$}
\psfrag{c}[b]{\hspace{0.79cm}  \scriptsize$y_\star=3$}
%\includegraphics[width=0.45\textwidth]{ctcbCRS} 
% \quad~
\psfrag{x}[]{\small $c_\ga^{\rm eff}$}
\psfrag{y}[b]{\small $c_g^{\rm eff}$}
\psfrag{a}[b]{\hspace{1cm}  \scriptsize $y_\star=0.5$}
\psfrag{b}[b]{\hspace{1.03cm}  \scriptsize $y_\star=1.5$}
\psfrag{c}[b]{\hspace{0.79cm}  \scriptsize$y_\star=3$}
\includegraphics[width=\textwidth]{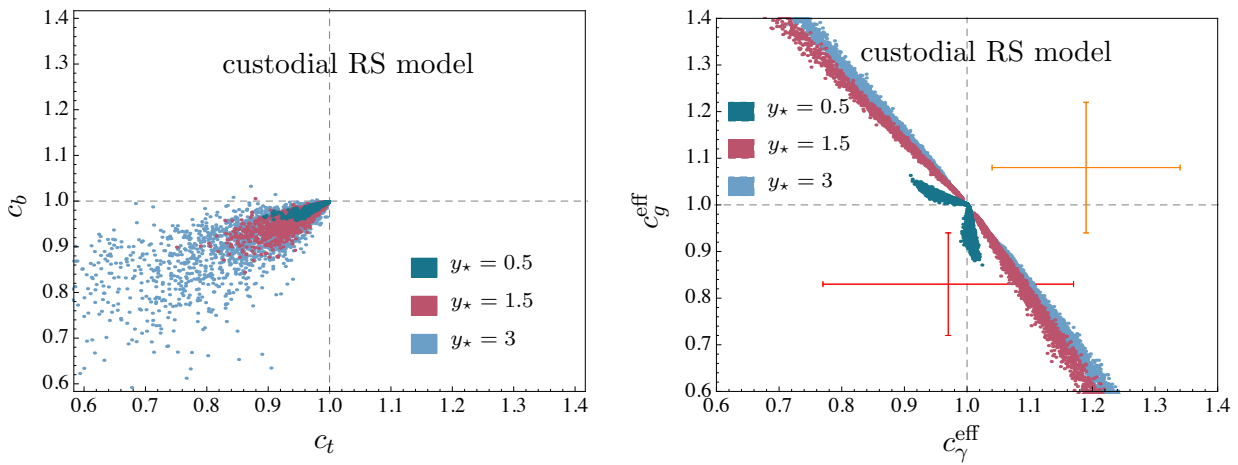} 
\parbox{15.5cm}
{\caption{\label{fig:kgkga}
Correlation between the Higgs couplings $c_t$ and $c_b$ (left) and the effective Higgs couplings $c_\ga^{\rm eff}$ and  $c_g^{\rm eff}$ (right) in the custodial RS model. All points obey the constraint $M_{g^{(1)}}>4.8$\,TeV imposed by a tree-level analysis of the $S$ and $T$ parameters at 95\% CL. In the right plot, the orange (red) cross represents the experimental values (with $1\sigma$ errors) obtained by ATLAS (CMS).}}
\end{center}
\end{figure}

We have explained earlier that, to good approximation, the average results for the various Higgs couplings in RS models can be expressed in terms of only two parameters $\mkk$ and $y_\star$, with some relatively narrow distribution of model points about these average predictions. As a result, in these models there are strong correlations between various Higgs couplings. This important fact is illustrated in Figure~\ref{fig:kgkga}, where we display our predictions in the $c_t$\,--\,$c_b$ and $c_\ga^{\rm eff}$\,--\,$c_g^{\rm eff}$ planes. In the right plot, scatter points below $c_g^{\rm eff}=1$ (lower right plane) correspond to the brane-localized Higgs scenario, while points above $c_g^{\rm eff}=1$ (upper left plane) refer to the narrow bulk-Higgs scenario. All points included in these plots obey the constraint $M_{g^{(1)}}>4.8$\,TeV implied by electroweak precision tests. In the case of the fermionic couplings $c_t$ and $c_b$ we observe a clear correlation in the sense that both couplings are smaller than~1 by approximately equal amounts. On the other hand, we see a clear anti-correlation between $c_\ga^{\rm eff}$ and $c_g^{\rm eff}$, which is due to the fermion KK contributions as explained above. This implies that there are no regions of parameter space where both couplings are smaller or larger than~1. Thus, a precise measurement of such values could rule out all RS scenarios considered in this work. The orange and red crosses in the right plot indicate the $1\sigma$ fit values $c_g^{\rm eff, exp}=1.08^{+0.15}_{-0.13}$, $c_\ga^{\rm eff, exp}=1.19^{+0.15}_{-0.12}$ \cite{ATLASMoriond2014} and $c_g^{\rm eff, exp}=  0.83^{+0.11}_{-0.10}$, $c_\ga^{\rm eff, exp}=0.97^{+0.17}_{-0.20}$ \cite{CMSMoriond2013} reported by the ATLAS and CMS Collaborations. Those fit values have a slight tendency to values larger (smaller) than 1 for both couplings in case of ATLAS (CMS), but they are compatible with our predictions within the error bars. Note that we have to be cautious when comparing our theoretical predictions with the fit values in question, because they have been obtained by varying $c_g^{\rm eff}$ and $c_\ga^{\rm eff}$ so as to obtain the best fit values to the experimental data {\em assuming\/} that the tree-level Higgs couplings take their SM values. It would be much preferable -- and the clearest way to test any new-physics model -- to compare the theoretical predictions with future results from {\em model-independent\/} analyses of the Higgs couplings.

\subsubsection*{Future sensitivities on Higgs couplings of LHC and ILC}

In the last part of this section, we wish to illustrate the potential for constraining the relevant parameters of the RS models by future, model-independent analyses of Higgs couplings. It has been reported in \cite{Peskin:2012we} that the LHC with an integrated luminosity of $300\,\rm{fb}^{-1}$ has the potential to probe deviations of Higgs couplings to fermions in the range of 14\%\,--\,46\% and to gauge bosons in the range of 14\%\,--\,30\%, both at 95\% CL. At future lepton colliders like the ILC \cite{Baer:2013cma,Klute:2013cx,Asner:2013psa,Tian:2013yda} the sensitivity to deviations can be improved by almost one order of magnitude. In the following analysis we focus on the LHC operating at $\sqrt{s}=14$\,TeV with $300\,\rm{fb}^{-1}$ of integrated luminosity and the ILC  operating at $\sqrt{s}=1$\,TeV with integrated luminosity of $1000\,\rm{fb}^{-1}$.

\begin{figure}[t]
\begin{center}
\psfrag{x}[]{\small $M_{g^{(1)}}~{\rm [TeV]}$}
\psfrag{w}[r]{\footnotesize $c_W$}
\psfrag{v}[r]{\footnotesize $c_Z$}
\psfrag{u}[r]{\footnotesize $c_t$\hspace{0mm}}
\psfrag{t}[r]{\footnotesize $c_b$\hspace{0mm}}
\psfrag{s}[r]{\footnotesize $c_\tau$\hspace{0mm}}
\psfrag{r}[r]{\footnotesize $c_g^{\rm eff}$ \rm (b.)}
\psfrag{q}[r]{\footnotesize $c_\ga^{\rm eff}$  \rm (b.)}
\psfrag{p}[r]{\footnotesize $c_g^{\rm eff}$  \rm (n.b.)\hspace{0mm}}
\psfrag{o}[r]{\footnotesize $c_\ga^{\rm eff}$  \rm (n.b.)\hspace{0mm}}
\psfrag{n}[r]{\footnotesize $c_t$\hspace{0mm}}
\psfrag{m}[r]{\footnotesize $c_b$\hspace{0mm}}
\psfrag{l}[r]{\footnotesize $c_\tau$\hspace{0mm}}
\psfrag{k}[r]{\footnotesize $c_g^{\rm eff}$ \rm (b.)}
\psfrag{j}[r]{\footnotesize $c_\ga^{\rm eff}$  \rm (b.)}
\psfrag{i}[r]{\footnotesize $c_g^{\rm eff}$  \rm (n.b.)\hspace{0mm}}
\psfrag{h}[r]{\footnotesize $c_\ga^{\rm eff}$  \rm (n.b.)\hspace{0mm}}
\psfrag{y}[b]{\small LHC ($14\,\rm TeV$, $300\, \rm fb^{-1}$)}
\psfrag{z}[b]{\small ILC ($1\,\rm TeV$, $1000\, \rm fb^{-1}$)}
\psfrag{a}[left]{\small \hspace{-5mm} $y_\star=3$}
\psfrag{b}[left]{\small \hspace{-5mm} $y_\star=1.5$}
\includegraphics[width=\textwidth]{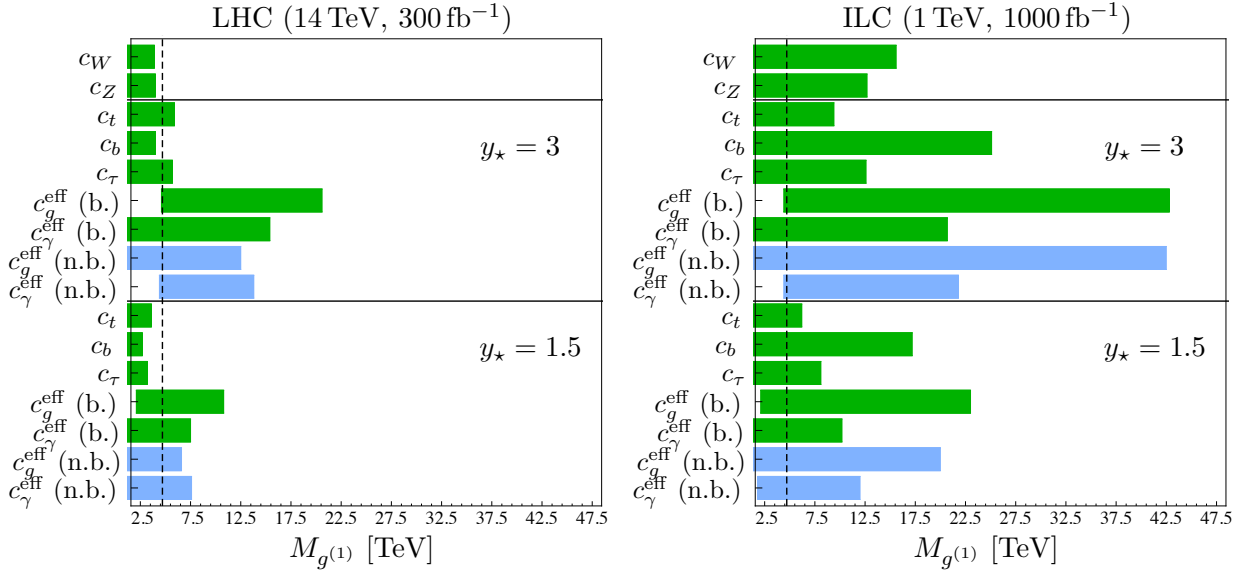}
\parbox{15.5cm}
{\caption{\label{fig:kbounds}
Summary of the exclusion limits (at 95\% CL) on the mass of the first KK gluon resonance in the custodial RS model, which could be derived from SM-like measurements of Higgs couplings at the high-luminosity LHC (left) and the ILC (right), for two representative values of $y_\star$. For the loop-induced couplings $c_g^{\rm eff}$ and $c_\ga^{\rm eff}$, we distinguish between the brane (green) and the narrow bulk-Higgs (blue) scenarios. The dashed vertical lines show the lower bounds on $M_{g^{(1)}}$ obtained from electroweak precision measurements.}}
\end{center}
\end{figure}

\begin{table}[ht]
\begin{center}
\begin{tabular}{|c|cccc|} 
\hline
$c_i^{\rm (eff)}-1$ & $W$ & $Z$ & $g$ & $\ga$ \\ 
\hline
LHC $14\,\TeV, 300\,\rm{fb}^{-1}$ & $(-0.069,0)$ & $(-0.077,0)$ & $(-0.078,0.10)$ 
 & $\!(-0.096,0.059)\!$ \\
ILC $1\,\TeV, 1000\,\rm{fb}^{-1}$ & $(-0.004,0)$ & $(-0.006,0)$ & $(-0.014,0.014)$ 
 & $\!(-0.032,0.035)\!$ \\
\hline
$c_i-1$ & $t$ & $b$ & $\tau$ & \\ 
\hline
LHC $14\,\TeV, 300\,\rm{fb}^{-1}$ & $(-0.154,0.147)$
 & $(-0.231,0.041)$ & $(-0.093,0.132)$ & \\
ILC $1\,\TeV, 1000\,\rm{fb}^{-1}$ & $(-0.044,0.035)$
 & $(-0.003,0.011)$ & $(-0.013,0.017)$ & \\
\hline
\end{tabular}
\parbox{15.5cm}
{\caption{\label{tab:kexp}
Experimental capabilities for model-independent measurements of the Higgs-boson couplings $c_i$ to gauge bosons (top) and third-generation fermions (bottom), expressed as $1\sigma$ confidence intervals derived in \cite{Peskin:2012we}. For the case of the $hgg$ and $h\ga\ga$ couplings we show the effective coefficients $c_{g,\ga}^{\rm eff}$ defined in (\ref{eqn:cgeff}) and (\ref{eqn:cga5}).}}
\end{center}
\end{table} 

Our goal is to derive exclusion bounds for the mass of the first KK gluon resonance from each of the Higgs couplings. To obtain these bounds, we plot each coupling $c_i$ as in Figure~\ref{fig:ctct5}, fit a Gaussian distribution to the model points for each pair of $y_\star$ and $M_{g^{(1)}}$, and determine the mean values $c_i$ with the standard deviations $\sigma_{c_i}$. For the experimental couplings we assume that they are SM-like, $c_i^{\rm exp}=1$, with the $1\sigma$ errors given in Table~\ref{tab:kexp}. These errors are asymmetric and correspond to the $1\sigma$ confidence intervals as they emerge from the combined fit (subject to certain assumptions) performed in \cite{Peskin:2012we}. We then consider the ratio $c_i/c_i^\tx{exp}=c_i$, and calculate the corresponding standard deviation by combining the theoretical and experimental errors in quadrature. Finally, we test at which confidence level the coefficient $c_i$ is compatible with~1. The results are compiled in Figure~\ref{fig:kbounds} for two representative values of $y_\star$. The colored regions are the 95\% CL excluded regions for the mass of the lightest KK gluon resonance. To obtain exclusion bounds for arbitrary values of $y_\star$, one can make use of the fact that the exclusion limits depend linearly on $y_\star$ to good approximation. We see that the strongest bounds emerge from the loop-induced Higgs couplings, for which we distinguish between the brane-Higgs (b.)~and narrow bulk-Higgs (n.b.) scenarios. Our results imply that the high-luminosity run at the LHC can probe or exclude KK gluon masses in the range $M_{g^{(1)}}<21\,{\rm TeV}\times (y_\star/3)$ for the brane Higgs scenario, and $M_{g^{(1)}}<13\,{\rm TeV}\times (y_\star/3)$ in the narrow bulk-Higgs model. For the ILC, one expects to probe or rule out KK gluon masses in the range $M_{g^{(1)}}<43\,{\rm TeV}\times (y_\star/3)$ in both scenarios.\footnote{The different limits in the case of the LHC are due to the asymmetric error margins for $c_g$, see Table~\ref{tab:kexp}.} 
Note also that, independently of the realization of the Yukawa sector (and hence the parameter $y_\star$), the analysis of the Higgs couplings to $W$ bosons at the ILC is expected to be sensitive to KK gluon masses of up to 15\,TeV. In all cases, these limits by far exceed the mass ranges allowing for a direct discovery of KK resonances.

\section{Analysis of signal rates in the custodial RS model} 
\label{sec:bounds}

We finally investigate in more detail the Higgs decay rates into pairs of electroweak gauge bosons and third-generation fermions. In order to directly compare our predictions with experimental measurements, we study the signal rates $R_X$ defined in \eqref{eqn:RXXintro}, which can be expressed in terms of the effective couplings $c_i$ and $c_{i5}$ derived in Section~\ref{sec:coupsA} via
\begin{equation}\label{eqn:RXX}
   R_X 
   \equiv \frac{(\sigma\cdot{\rm BR)}(pp\to h\to X)_\RS}%
               {(\sigma \cdot {\rm BR)}(pp\to h\to X)_\SM} 
   = \frac{\big[\big(|c_g^{\rm eff}|^2 + |c_{g5}^{\rm eff}|^2\big) f_{\rm GF}
           + c_V^2 f_{\rm VBF} \big] \big[|c_X^{\rm (eff)}|^2 
           + |c_{X5}^{\rm (eff)}|^2 \big]}{c_h} \,.
\end{equation}
The correction to the total Higgs width relative to the SM total width $\Gamma^\SM_h=4.14$\,MeV (for $m_h =125.5$\,GeV) can be accounted for by the parameter \cite{Denner:2011mq}
\begin{equation}\label{eqn:kh}
   c_h = \frac{\Gamma_h^\tx{RS}}{\Gamma_h^\tx{SM}}
   \approx 0.57 (c_b^2 + c_{b5}^2) + 0.22 c_W^2 + 0.03 c_Z^2
    + 0.09 \big(|c_g^{\rm eff}|^2 + |c_{g5}^{\rm eff}|^2\big) 
    + 0.06 (c_\tau^2 + c_{\tau5}^2) + 0.03 \,.
\end{equation}
The corrections to the decay modes $h\to c\bar c,\,Z\ga,\dots$ have a numerically insignificant effect and can therefore be neglected; the combined branching fraction of these modes is 3\% in the SM. In \eqref{eqn:RXX} we have taken into account the probabilities to produce a Higgs boson via gluon fusion (GF), or via vector-boson fusion and associated $hV$ production (collectively referred to as VBF). Concerning the latter production processes, we have implemented the findings of Section~\ref{sec:VBF}, showing that the leading  corrections proportional to $L$ to the corresponding cross sections are given by $c_V^2$, where in the custodial RS model there is no need to distinguish between $c_W$ and $c_Z$ as far as these terms are concerned, see (\ref{eqn:cWZcust}). Other production channels such as $pp\to ht\bar t$ can be neglected to very good approximation. For inclusive Higgs production at the LHC the appropriate fractions are $f_{\rm GF}\approx 0.9$ and $f_{\rm VBF}\approx 0.1$. For the case of the final state $X=b\bar b$, Higgs-strahlung is an experimentally more feasible Higgs production channel at the LHC than gluon fusion, since the latter suffers from an overwhelming QCD background \cite{ATLAS-CONF-2013-079}. For the case of the signal rate $R_{bb}$ we thus have to set $f_\tx{GF}=0$ and $f_\tx{VBF}=1$ in \eqref{eqn:RXX}. A further comment concerns the Higgs decays into $WW^*$ and $ZZ^*$, with subsequent decays of the off-shell vector boson into fermions. According to the discussion in Section~\ref{sec:hVV}, we use the expression for $\Gamma(h\to VV^*)/\Gamma(h\to VV^*)_{\rm SM}$ derived from \eqref{GhWWfin} instead of $c_V^2$ in this case.

\begin{table}[t]
\begin{center}
\begin{tabular}{|c|ccccc|} 
\hline
$R_X$ & $bb$ & $\tau\tau$ & $WW$ & $ZZ$ & $\ga\ga$ \\ 
\hline
ATLAS \cite{ATLASMoriond2014} & $0.2^{+0.7}_{-0.6}$ & $1.4^{+0.5}_{-0.4}$
 & $1.00^{+0.32}_{-0.29}$ & $1.44^{+0.40}_{-0.35}$ & $1.57^{+0.33}_{-0.28}$ \\
CMS \cite{CMSMoriond2013} & $1.0^{+0.5}_{-0.5}$ \cite{Chatrchyan:2013zna}
 & $0.78^{+0.27}_{-0.27}$ \cite{Chatrchyan:2014nva} & $0.68^{+0.20}_{-0.20}$
 & $0.92^{+0.28}_{-0.28}$ & $0.77^{+0.27}_{-0.27}$ \\
Average & $0.7^{+0.4}_{-0.4}$ & $0.92^{+0.24}_{-0.22}$ & $0.77^{+0.17}_{-0.16}$
 & $1.09^{+0.23}_{-0.22}$ & $1.09^{+0.21}_{-0.19}$ \\
\hline
\end{tabular}
\parbox{15.5cm}
{\caption{\label{tab:Rexp} 
Experimental values for the signal rates measured by the ATLAS and CMS Collaborations including the $1\sigma$ errors. The assumed Higgs masses are $m_h=125.5$\,GeV in \cite{ATLASMoriond2014}, $m_h=125.7$\,GeV in \cite{CMSMoriond2013}, and $m_h=125$\,GeV in \cite{Chatrchyan:2013zna,Chatrchyan:2014nva}.}}
\end{center}
\end{table}

In the following analysis we will focus first on the individual Higgs decay rates in the context of the custodial RS model. We will then present a summary of the bounds on the KK gluon mass $M_{g^{(1)}}$ and the parameter $y_\star$, which are derived by confronting our predictions with naive averages of the signal strengths reported by the ATLAS and CMS Collaborations and summarized in Table~\ref{tab:Rexp}. A more thorough analysis properly accounting for correlations between the various measurements should be performed by the experimental collaborations. 

\subsubsection*{\boldmath Analysis of the signal rates $R_{\gamma\gamma}$, $R_{ZZ}$, and $R_{WW}$}

\begin{figure}[t]
\begin{center}
\psfrag{x}[]{\small $M_{g^{(1)}}~{\rm [TeV]}$}
\psfrag{y}[b]{\small $R_{ZZ}$}
\psfrag{a}[b]{\hspace{1cm} \scriptsize $y_\star=0.5$}
\psfrag{b}[b]{\hspace{1.03cm} \scriptsize $y_\star=1.5$}
\psfrag{c}[b]{\hspace{0.79cm} \scriptsize$y_\star=3$}
\psfrag{z}[]{\small \vspace{2cm} \begin{tabular}{l} custodial RS model \\[-1mm] narrow bulk Higgs \end{tabular}}
\psfrag{w}[]{\small\hspace{4mm} \begin{tabular}{l} custodial RS model \\[-1mm] brane Higgs \end{tabular}}
\includegraphics[width=\textwidth]{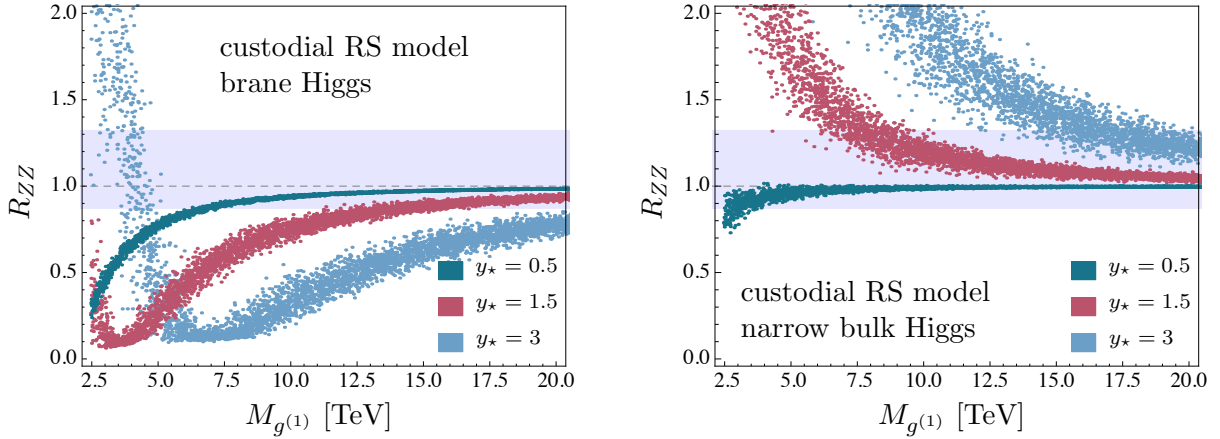}
\parbox{15.5cm}
{\caption{\label{fig:RZZCRS} 
Predictions for the ratio $R_{ZZ}$ as a function of the KK gluon mass $M_{g^{(1)}}$ in the custodial RS model, for the cases of a brane-localized Higgs boson (left) and a narrow bulk-Higgs field (right). The scatter points with different color correspond to different values of $y_\star$. The blue band represents the $1\sigma$ experimental error range for the observable $R_{ZZ}$.}}
\end{center}
\end{figure}

We start our analysis with a discussion of Higgs decays into two electroweak gauge bosons. The decay into two photons has been discussed extensively in our previous work \cite{Hahn:2013nza}, see in particular Figure~4 in this reference. We will not repeat the corresponding analysis here. Figure~\ref{fig:RZZCRS} shows the results for the ratio $R_{ZZ}$ as a function of the mass $M_{g^{(1)}}$ of the lightest KK gluon state and for three different values for $y_\star$.\footnote{The process $pp\to h\to ZZ^*$ was also considered in our work \cite{Malm:2013jia}, where we did not take into account the Higgs production process via vector-boson fusion as well as the modifications of the total Higgs width and the $h\to ZZ^*$ decay rate. Consequently, the analysis presented here is more accurate.} 
To excellent approximation the scatter points also represent the results for the observable  $R_{WW}$, since at the level of the $L$-enhanced terms the Higgs decays into $ZZ^*$ and $WW^*$ are expressed by the same modification factor $c_Z^2\approx c_W^2$, see \eqref{GhWWfin} and \eqref{eqn:cWZcust}. The blue band represents the $1\sigma$ error range corresponding to the latest experimental values for $R_{ZZ}$ given in Table~\ref{tab:Rexp}, where the naively averaged value has been used. Model points falling outside this band are excluded at 68\% CL. (Alternatively we could have used the average experimental value for the ratio $R_{WW}$, in which case the excluded set of model points is a different one.) It is interesting to observe that for relatively large values for $y_\star$ the data already disfavor KK gluon masses in the low TeV range. The tensions between the theoretical predictions for $R_{ZZ}$ ($R_{WW}$) and the experimental data are stronger for the brane-Higgs (narrow bulk-Higgs) model due to the mild tendency of an enhanced (suppressed) cross section seen in the data, which is in conflict with the suppression (enhancement) of the predicted cross section. 
 
The shapes of the curves can be explained by the fact that, for not too small Yukawa couplings, the RS corrections to the gluon-fusion cross section by far dominate over the corrections to the Higgs decay rates. The results then closely resemble those shown in Figure~5 of \cite{Malm:2013jia}, where only the corrections to the gluon-fusion cross section were taken into account. The dependance of this production channel on the details of the localization of the Higgs profile on or near the IR brane explains why the ratios $R_{VV}$ are suppressed (enhanced) in the brane-localized (narrow bulk-Higgs) scenario. For small values of $M_{g^{(1)}}$ and $y_\star$, however, the loop-induced couplings become subdominant, and the negative corrections to the $h\to ZZ^*$ decay width give rise to a reduction of the signal rate even in the narrow bulk-Higgs scenario. The peculiar behavior seen for very small KK scales in the left plot in Figure~\ref{fig:RZZCRS} can be understood as follows. For $y_\star=3$, the $gg\to h$ production cross section vanishes for $M_{g^{(1)}}\approx 7.0$\,TeV, because the new-physics contribution cancels the SM amplitude. However, due to the vector-boson fusion production process a non-zero value of $R_{ZZ}$ remains. For even smaller values of $M_{g^{(1)}}$ the new-physics amplitude dominates over the SM one and the cross section rises again.

\begin{figure}[t]
\begin{center}
\psfrag{x}[]{$R_{\gamma \gamma}$}
\psfrag{y}[b]{$R_{ZZ}$}
\psfrag{a}[bl]{\footnotesize $y_\star=0.5$}
\psfrag{b}[bl]{\footnotesize$y_\star=1.5$}
\psfrag{c}[bl]{\footnotesize$y_\star=3$}
\psfrag{z}[]{}
\psfrag{w}[]{}
\begin{tikzpicture}[scale=1]
\draw (0,0) node {\includegraphics[width=0.55\textwidth]{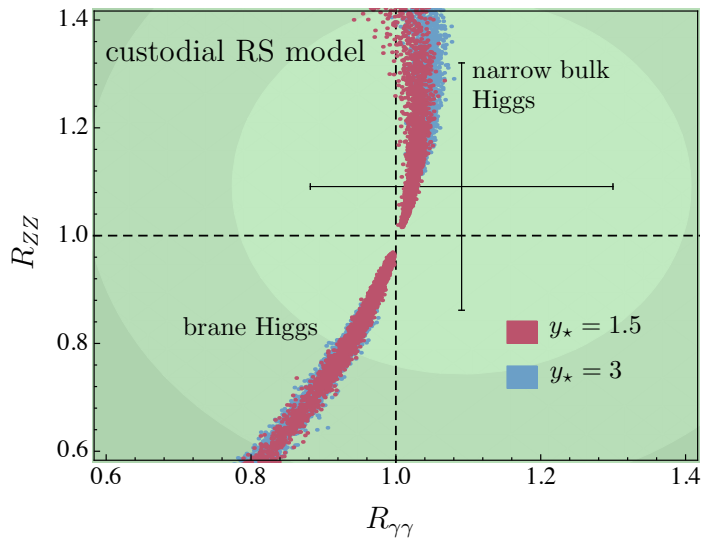}};
%\draw (-1.7,2.8) node {\rm custodial RS model};
%\draw (-1.5,-1.0) node {\footnotesize \rm brane Higgs};
%\draw (2.5,2.3) node {\footnotesize\rm \parbox{2cm}{narrow bulk \\ Higgs}};
\end{tikzpicture}
\parbox{15.5cm}
{\caption{\label{fig:RgagaRZZ}
Correlation of the predictions for the signal rates $R_{ZZ}$ and $R_{\ga\ga}$ in the custodial RS model under variations of $M_{g^{(1)}}$ and $y_\star$. All scatter points fulfill the constraints from electroweak precision tests. The cross shows the average experimental values with $1\sigma$ errors for the measured signal rates.}}
\end{center}
\end{figure}

The new-physics effects on the ratios $R_{ZZ}$ and $R_{WW}$ are stronger than those on $R_{\ga\ga}$, since in the latter case there is a partial compensation between the contributions of fermionic KK resonances to the Higgs production cross section via gluon fusion and to the $h\to\ga\ga$ decay rate \cite{Hahn:2013nza}. The strong correlation between $R_{ZZ}$ and $R_{\ga\ga}$ resulting from these fermionic corrections is examined in Figure~\ref{fig:RgagaRZZ}. The SM predicts the values $R^{\rm SM}_{ZZ,\ga\ga}=1$ denoted by the crossing position of the dashed lines. Scatter points below the horizontal dashed line belong to the brane-localized Higgs scenario, while the points above the line belong to the narrow bulk-Higgs scenario.\footnote{We only show scatter points for $y_\star=1.5$ and 3. For $y_\star=0.5$, both $R_{ZZ}$ and $R_{\gamma\gamma}$ are always reduced, see Figure~\ref{fig:RZZCRS} and Figure~4 in \cite{Hahn:2013nza}.} 
All scatter points fulfill the bound $M_{g^{(1)}}>4.8$\,GeV imposed by the measurements of the oblique parameters $S$ and $T$. The cross shows the experimental values given in Table~\ref{tab:Rexp}, while the green ellipses present the 68\%, 95\%, and 99\% confidence regions of the combined measurements. We observe a strong correlation between the two ratios, where for reasons explained above the new-physics effects are larger for $R_{ZZ}$ than for $R_{\ga\ga}$. Notice that the naively averaged current experimental data slightly favor the narrow bulk-Higgs over the brane-localized Higgs scenario. It should be noted, however, that using only the measurements reported by CMS one would obtain the opposite conclusion. 

\subsubsection*{\boldmath Analysis of the signal rates $R_{\tau\tau}$, $R_{bb}$ and the total Higgs width}

\begin{figure}[t]
\begin{center}
\psfrag{x}[]{\small $M_{g^{(1)}}~{\rm [TeV]}$}
\psfrag{y}[b]{\small $R_{\tau\tau}$}
\psfrag{a}[b]{\hspace{1cm} \scriptsize $y_\star=0.5$}
\psfrag{b}[b]{\hspace{1.03cm} \scriptsize $y_\star=1.5$}
\psfrag{c}[b]{\hspace{0.79cm} \scriptsize $y_\star=3$}
\psfrag{z}[]{\small \hspace{2mm} \begin{tabular}{l} custodial RS model \\[-1mm] narrow bulk Higgs \end{tabular}}
\psfrag{w}[]{\small\hspace{4mm} \begin{tabular}{l} custodial RS model \\[-1mm] brane Higgs \end{tabular}}
\includegraphics[width=\textwidth]{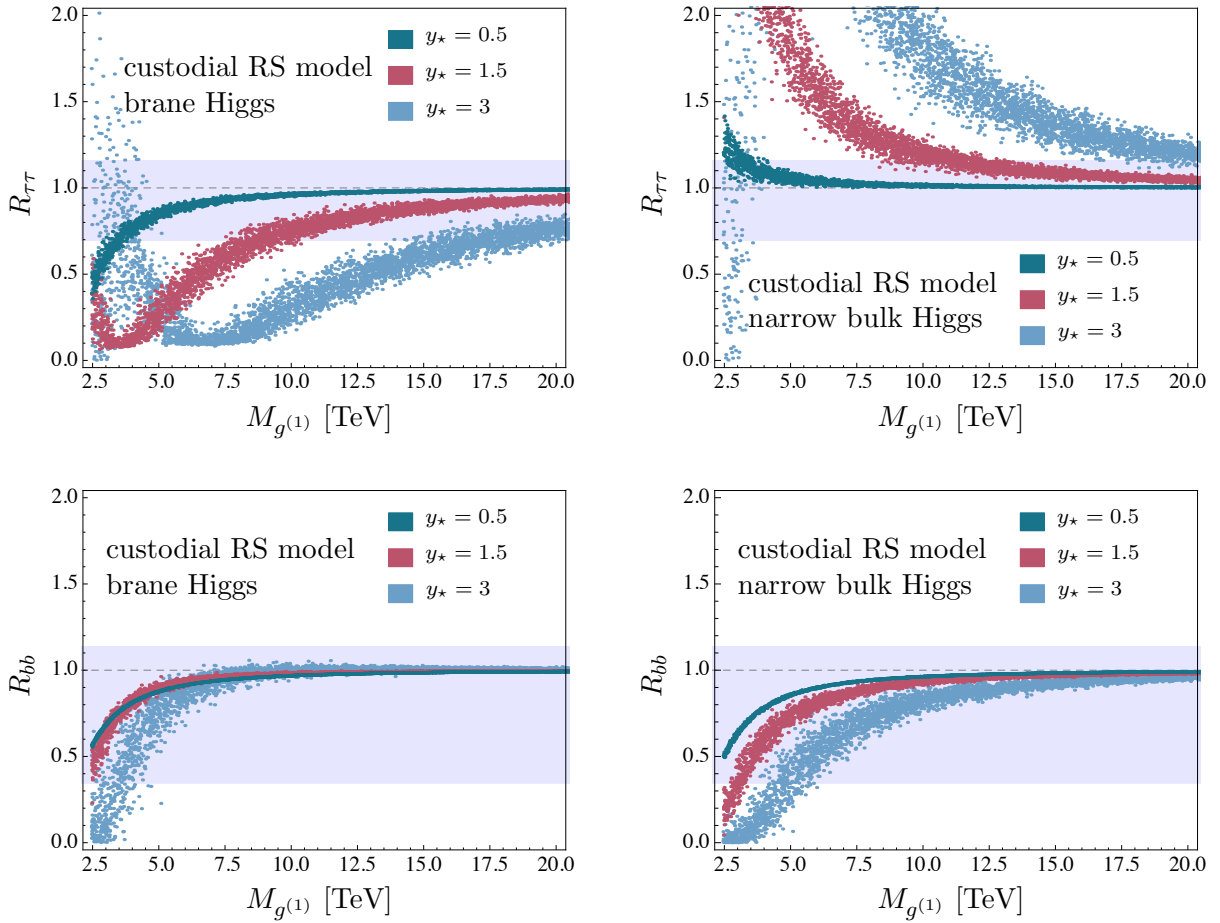}
\psfrag{x}[]{\small $M_{g^{(1)}}~{\rm [TeV]}$}
\psfrag{y}[b]{\small $R_{bb}$}
\psfrag{a}[b]{\hspace{1cm} \scriptsize $y_\star=0.5$}
\psfrag{b}[b]{\hspace{1.03cm} \scriptsize $y_\star=1.5$}
\psfrag{c}[b]{\hspace{0.79cm} \scriptsize$y_\star=3$}
\psfrag{z}[]{\small \hspace{12mm} \begin{tabular}{l} custodial RS model \\[-1mm]  brane Higgs  \end{tabular}}
\psfrag{w}[]{\small\hspace{12mm} \begin{tabular}{l} custodial RS model \\[-1mm] narrow bulk Higgs\end{tabular}}
\parbox{15.5cm}
{\caption{\label{fig:RtautauCRS} 
Predictions for the ratios $R_{\tau\tau}$ (upper plots) and $R_{bb}$ (lower plots) as a function of the KK gluon mass $M_{g^{(1)}}$ in the custodial RS model, for the cases of a brane-localized Higgs boson (left) and a narrow bulk-Higgs field (right). The meaning of the colors is the same as in Figure~\ref{fig:ctct5}.}}
\end{center}
\end{figure}
 
We now turn to the predictions for $R_{\tau\tau}$ and $R_{bb}$ in the custodial RS model. The upper plots in Figure~\ref{fig:RtautauCRS} show the observable $R_{\tau\tau}$ as a function of $M_{g^{(1)}}$. As in the previous cases, the shapes of the curves are largely due to the behavior of the Higgs-boson production cross section, which is dominated by the gluon-fusion process. Particularly for small KK scales, these effects are quite large and have the potential to compensate and even exceed the SM contribution. For very small KK scales ($M_{g^{(1)}}\lesssim 3$\,TeV), on the other hand, the negative corrections to the $c_\tau$ coupling can become so large that the $h\to\tau^+\tau^-$ decay rate almost vanishes (see Figure~\ref{fig:ctct5}), and hence $R_{\tau\tau}$ can drop close to zero. The observable $R_{bb}$ shown in the lower plots receives more moderate corrections, since in this case the only production channel included is Higgs-strahlung. Although there is no need to distinguish between the brane-localized and narrow bulk-Higgs scenario in the Higgs production cross section and the $h\to b\bar b$ decay rate, the two plots still differ due to the contribution of the $h\to gg$ decay rate to the total Higgs width. This partial rate is reduced in the brane-Higgs scenario and enhanced in bulk-Higgs models. 

\begin{figure}[t]
\begin{center}
\psfrag{x}[]{\small $M_{g^{(1)}}~{\rm [TeV]}$}
\psfrag{y}[b]{\small $c_{h}$}
\psfrag{a}[b]{\hspace{1cm} \scriptsize $y_\star=0.5$}
\psfrag{b}[b]{\hspace{1.03cm} \scriptsize $y_\star=1.5$}
\psfrag{c}[b]{\hspace{0.79cm} \scriptsize$y_\star=3$}
\psfrag{z}[]{\small \hspace{2mm} \begin{tabular}{l} custodial RS model \\[-1mm] narrow bulk Higgs \end{tabular}}
\psfrag{w}[]{\small\hspace{4mm} \begin{tabular}{l} custodial RS model \\[-1mm] brane Higgs \end{tabular}}
\includegraphics[width=0.975\textwidth]{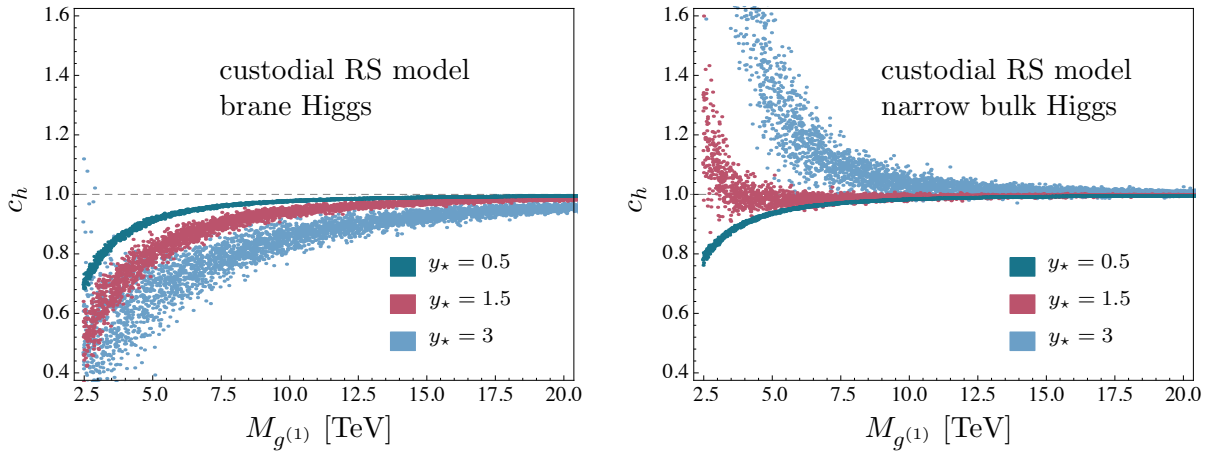}
\parbox{15.5cm}
{\caption{\label{fig:Hwidth} 
Predictions for $c_h=\Gamma_h^{\RS}/\Gamma_h^\SM$ as a function of $M_{g^{(1)}}$ in the custodial RS model, for the cases of a brane-localized (left) and a narrow bulk-Higgs field (right).}}
\end{center}
\end{figure}

The present data on $R_{bb}$ only imply weak constraints on the RS parameter space, because the experimental accuracy is worse than for all other channels, see Table~\ref{tab:Rexp}. Nevertheless, the Higgs coupling to bottom quarks $c_b$ is an important quantity, since it gives rise to one of the most significant corrections to the total Higgs width \eqref{eqn:kh}, which enters all of the signal rates in (\ref{eqn:RXXintro}). Figure~\ref{fig:Hwidth} shows the ratio $c_h=\Gamma_h^{\RS}/\Gamma_h^\SM$ in the custodial RS model. We see that in the brane-Higgs scenario the Higgs width can be reduced by about 25\,--\,50\% (10\,--\,20\%) for a KK gluon mass $M_{g^{(1)}}\approx 5\,\TeV\,\, (10\,\TeV)$ and maximal Yukawa value $y_\star=3$. The dominant effects come from the decays $h\to b\bar b$ and $h\to gg$, both of which receive negative corrections. The situation is different in the case of the narrow bulk-Higgs scenario, where the $h\to gg$ decay rate receives a large positive correction, which enhances the Higgs width and counteracts the suppression of the $h\to b\bar b$ decay rate. This effect dominates for $y_\star\gtrsim 1.5$, leading to a Higgs width larger than in the SM. 

\subsubsection*{Summary of exclusion bounds for $\bs{M_{g^{(1)}}}$ and $\bs{y_\star}$}

\begin{figure}[t]
\begin{center}
\psfrag{x}[]{\small $M_{g^{(1)}}~{\rm [TeV]}$}
\psfrag{z}[]{$y_\star$}
\psfrag{y}[b]{}
\psfrag{a}[]{\footnotesize $y_\star=3$}
\psfrag{b}[]{\footnotesize $y_\star=1.5$}
\psfrag{w}[b]{\scriptsize $h\to b\bar b$\hspace{1cm} }
\psfrag{v}[b]{\scriptsize $h\to \tau^+\tau^-$\hspace{1.2cm}}
\psfrag{u}[b]{\scriptsize $h\to WW^*$\hspace{1.2cm}}
\psfrag{t}[b]{\scriptsize $h\to ZZ^*$\hspace{1cm}}
\psfrag{s}[b]{\scriptsize $h\to \gamma\gamma$\hspace{1cm}}
\psfrag{r}[b]{\scriptsize $h\to b\bar b$\hspace{1cm}}
\psfrag{q}[b]{\scriptsize $h\to \tau^+\tau^-$\hspace{1.2cm}}
\psfrag{p}[b]{\scriptsize $h\to WW^*$\hspace{1.2cm}}
\psfrag{o}[b]{\scriptsize $h\to ZZ^*$\hspace{1cm}}
\psfrag{n}[b]{\scriptsize $h\to \gamma\gamma$\hspace{1cm}}
\psfrag{c}[]{\footnotesize \hspace{1.4cm} $M_{g^{(1)}}=4.8$\,TeV}
\psfrag{d}[]{\footnotesize \hspace{1.3cm} $M_{g^{(1)}}=10$\,TeV}
\includegraphics[width=0.975\textwidth]{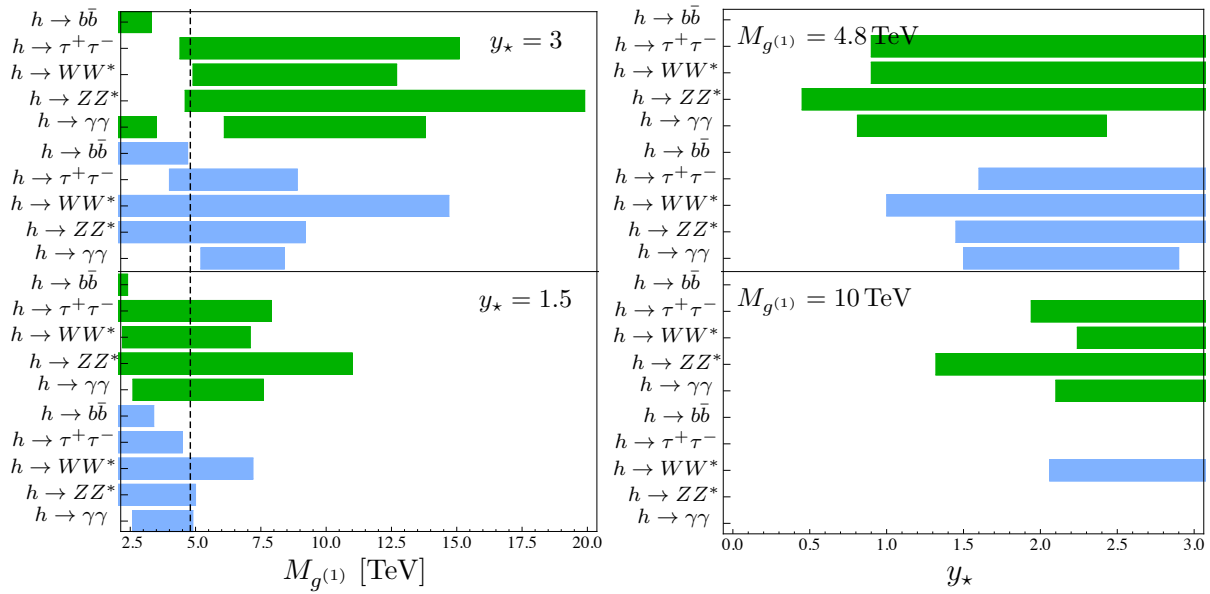}
\parbox{15.8cm}
{\caption{\label{fig:RboundCRS}
Summary of the bounds on the mass of the lightest KK gluon (left) and the parameter $y_\star$ (right) obtained from the exclusion plots in the custodial RS model for the brane-localized (green) and narrow bulk-Higgs scenario (blue). The shaded regions are excluded at $95\%$ CL for each corresponding decay channel. The vertical dashed line shows the bound obtained from a tree-level analysis of electroweak precision observables.}}
\end{center}
\end{figure}

Even at the present level of precision, the existing measurements of the signal rates for the various Higgs-boson decays provide strong constraints on the parameter space of the RS models under consideration. In Figure~\ref{fig:RboundCRS} we show the exclusion limits obtained at $95\%$ CL on the mass of the first KK gluon resonance and the maximum value $y_\star$ of the elements of the anarchic 5D Yukawa matrices in the custodial RS model, derived by an analysis of the various decay rates using the latest experimental results shown in Table~\ref{tab:Rexp}. To obtain these limits, we have fitted a Gaussian distribution to the model points for each pair of $M_{g^{(1)}}$ and $y_\star$, and determined the mean values $R_X^{\rm th}$ and the standard deviations $\sigma_{R_X^{\rm th}}$ for these parameters, in analogy with our treatment of the effective Higgs couplings in Section~\ref{sec:coupsA}. We have then calculated the ratios $R_X/R_X^{\rm exp}$, combined the theoretical and experimental errors in quadrature, and tested at which confidence levels these ratios are compatible with~1. The green (blue) bars in the figure refer to the brane-localized (narrow bulk-Higgs) RS scenario. The most stringent bounds emerge from the signal rates for $pp\to h\to ZZ^*, WW^*$. The former yields tighter constraints in the brane-localized Higgs scenario and the latter in the narrow bulk-Higgs scenario. 

Taking the most stringent bounds from Figure~\ref{fig:RboundCRS}, which are obtained for $y_\star=3$, we derive at 95\% CL the lower bounds
\begin{equation}\label{eqn:CRSbounds1}
   M_{g^{(1)}}\big|^\tx{custodial RS}_\tx{brane Higgs} > 19.9\,{\rm TeV}
    \quad \tx{and} \quad
   M_{g^{(1)}}\big|^\tx{custodial RS}_\tx{narrow bulk Higgs} > 14.9\,{\rm TeV}
\end{equation}
on the mass of the first KK gluon resonance. We stress that, since these bounds derived from Higgs physics are much stronger than those stemming from electroweak precision measurements, from a phenomenological point of view there is not much gained by implementing the custodial protection mechanism. While this mechanism can tame the large tree-level effects on the $T$ parameter and the $Zb\bar b$ couplings in RS models, we still find very large contributions to loop-induced processes in the Higgs sector. A similar observation has been made in the context of loop-induced flavor-changing neutral current processes such as $b\to s\gamma$ \cite{Blanke:2012tv}. However, the effects found here are far more pronounced. A possible way out (aside from gauge-Higgs unification models \cite{Contino:2003ve, Agashe:2004rs}, where the Higgs is identified with the fifth component of a 5D gauge field) is to lower $y_\star$. The right plot in Figure~\ref{fig:RboundCRS} summarizes the exclusion regions on $y_\star$ obtained for two different values of the lightest KK gluon mass. The analysis has been restricted to values for $y_\star$ below the perturbativity bound $y_\star\le y_{\rm max}\approx 3$ \cite{Csaki:2008zd,Malm:2013jia}. Again, the most stringent bounds come from the processes $pp\to h\to ZZ^*, WW^*$ and can be combined to give the constraints (at $95\%$ CL)
\begin{equation}
   y_\star\big|^\tx{custodial RS}_\tx{brane Higgs} < 0.4
    \quad \tx{and} \quad 
   y_\star\big|^\tx{custodial RS}_\tx{narrow bulk Higgs} < 1.1 \,,
\end{equation}
valid for $M_{g^{(1)}}=4.8$\,TeV. We see that in particular in the brane-Higgs scenario small values are preferred. However, too small Yukawa couplings would give rise to enhanced corrections to $\epsilon_K$ \cite{Csaki:2008zd} and hence they would reinforce the RS flavor problem. Also, for $y_\star<1$ it becomes difficult to reproduce the physical value of the top-quark mass.

\section{Conclusions} 
\label{sec:conclusions}

The discovery of a Higgs boson at the LHC \cite{ATLAS:2012gk,CMS:2012gu} has initiated a new era in elementary particle physics. The couplings of this new particle are found to be close to those predicted for the scalar boson of the SM. An explanation to the hierarchy problem is thus more urgently needed than ever. Precise measurements of the Higgs couplings to SM fermions and bosons provide an important tool for the discovery and the distinction of new-physics models addressing the hierarchy problem. In this paper, we have presented a comprehensive discussion of the effective Higgs couplings and all relevant signal rates for the production and decay of the Higgs boson at the LHC in the context of warped extra-dimension models with the scalar sector localized on or near the IR brane. 

For the first time, we have presented a thorough study of all new-physics effects in RS models on the decay rates for the processes $h\to VV^*$ (with $V=W,Z$), with the subsequent decay of the off-shell gauge boson into a fermion pair. We have also studied the new-physics effects on the Higgs-strahlung and vector-boson fusion production processes and shown that to very good approximation they can be accounted for by the corrections to the on-shell $hVV$ couplings $c_V$. This analysis has included the effects of virtual KK gauge bosons, which have been shown to be subleading (in $L$) with respect to the contributions stemming from the modified $hVV$ couplings. 

We have then summarized the expressions for the effective Higgs couplings to pairs of gauge bosons and fermions obtained within the context of warped extra-dimension models with the Higgs sector localized on or near the IR brane. The distinction between brane-Higgs and narrow bulk-Higgs scenarios becomes relevant for the contribution of fermionic KK resonances to the loop-induced Higgs couplings to photons and gluons. The corrections to the $hWW$ and $hZZ$ couplings are universal and given by the very simple formula (\ref{cWcZres}), which shows that corrections of more than a few percent can only be reached for KK masses close to the bound $M_{g^{(1)}}>4.8$\,TeV implied by electroweak precision tests. The corrections to the Higgs couplings to fermions scale like $\sim y_\star^2\,v^2/\mkk^2$ and can be significant for $M_{g^{(1)}}\lesssim 10$\,TeV and not too small values of $y_\star$. Even larger corrections can appear in the loop-induced Higgs couplings to gluons and photons, due to the high multiplicity of virtual KK particles propagating in the loop. The corresponding contributions to $c_g^{\rm eff}$ and $c_\ga^{\rm eff}$ are strongly anti-correlated. For instance, for $y_\star=3$ and a KK gluon mass $M_{g^{(1)}}=10$\,TeV, the relevant couplings in the custodial RS model with a narrow bulk Higgs are $c_g^{\rm eff}\approx 1.5$ and $c_\ga^{\rm eff}\approx 0.7$. Our analysis has included both the CP-even and CP-odd Higgs couplings. The CP-odd couplings to fermions can receive significant contributions from the 5D Yukawa couplings, while the CP-odd couplings to massive gauge bosons vanish. Concerning the loop-induced couplings to gluons and photons, the KK tower only contributes to the CP-even couplings, while the top-quark loop induces a contribution to the CP-odd couplings $c_{g5}^{\rm eff}$ and $c_{\gamma5}^{\rm eff}$. This gives rise to a potentially important contribution to the electric dipole moment of the electron, which can naturally be at the present level of sensitivity. 

In order elucidate the potential of future measurements at high-luminosity proton and lepton colliders to indirectly search for hints of a warped extra dimension, we have compared the predicted new-physics effects on the relevant couplings with the sensitivities that can be reached at the LHC with $\sqrt{s}=14$\,TeV and an integrated luminosity of $300\,\rm fb^{-1}$, and at the ILC with $\sqrt{s}=1$\,TeV and an integrated luminosity of $1000\,\rm fb^{-1}$. The exclusion bounds obtained in the RS model with custodial symmetry under the assumption of SM-like measurements are summarized in Figure~\ref{fig:kbounds}. At the ILC in particular, one will be able to probe KK gluon masses in the range over several tens of TeV from an analysis of the loop-induced Higgs couplings to gluons and photons. The analysis of the Higgs coupling to $W$ bosons at the ILC will have an expected sensitivity to KK gluon masses of $M_{g^{(1)}}\approx 15$\,TeV, which is independent of the realization of the Yukawa sector and hence the value of the parameter $y_\star$.

In the last section of the paper we have compared our predictions for the various Higgs signal rates with the latest data from the LHC. The strongest exclusion bounds originate from the Higgs decay rates into pairs of electroweak gauge bosons. In the custodial RS model, KK gluon masses lighter than $19.9\,{\rm TeV}\times (y_\star/3)$ in the brane-Higgs case and $14.9\,{\rm TeV}\times (y_\star/3)$ in the narrow bulk-Higgs scenario are excluded at 95\% CL. Our analysis shows that Higgs physics provides very sensitive probes of virtual effects from heavy KK excitations. Especially the signal rates for Higgs decays into pairs of electroweak gauge bosons, which primarily probe new-physics effects via the gluon-fusion production mechanism, could be used to either explain possible deviations in the corresponding cross sections or to derive strong bounds on the RS parameter space. These bounds are complementary to and often stronger than those from electroweak precision observables and rare flavor-changing processes. In the custodial RS model, the indirect effect of KK states on the Higgs-boson processes are strongly enhanced compared with the minimal model \cite{Malm:2013jia,Hahn:2013nza}, and hence the current experimental results on various Higgs decays already provide strong constraints. Even under the pessimistic assumption that the direct detection of KK resonances is out of reach at the LHC, one may still see sizable modifications of the $pp\to h\to X$ signal rates for $X=\ga\ga,\,ZZ^*,\,WW^*,\,\tau^+\tau^-$, even with $M_{g^{(1)}}$ as heavy as 10 or 15\,TeV. It will be exciting to compare our predictions with future, more precise experimental results. Even if no KK particles are to be discovered at the LHC, such an analysis could still provide a hint of the existence of a warped extra dimension.

\vspace{3mm}
{\em Acknowledgements:\/}
We are grateful to Martin Bauer, Juliane Hahn and Clara H\"orner for useful discussions. This research has been supported by the Advanced Grant EFT4LHC of the European Research Council (ERC), the Cluster of Excellence {\em Precision Physics, Fundamental Interactions and Structure of Matter\/} (PRISMA -- EXC 1098), grant 05H12UME of the German Federal Ministry for Education and Research (BMBF), and the DFG Graduate School GRK~1581 {\em Symmetry Breaking in Fundamental Interactions}.

\newpage
\begin{appendix}

\section{\boldmath $h\to VV^*$ couplings in the custodial RS model} 
\label{app:cust}

\renewcommand{\theequation}{A.\arabic{equation}}
\setcounter{equation}{0}

The motivation for the custodial RS model has been to mitigate the large corrections to electroweak precision observables encountered in the minimal version of the model, especially those to the $T$ parameter \cite{Agashe:2003zs,Csaki:2003zu} and the $Zb\bar b$ couplings \cite{Agashe:2006at}. In this way some of the lightest KK particles can be in reach for a direct detection at the LHC \cite{Carena:2006bn,Cacciapaglia:2006gp,Contino:2006qr}. The custodial protection is achieved by means of an enlarged gauge group in the bulk of the extra dimension. We focus on a model with the bulk gauge symmetry $SU(3)_c\times SU(2)_L\times SU(2)_R\times U(1)_X\times P_{LR}$, where the two $SU(2)$ groups are broken down to the vectorial $SU(2)_V$ on the IR brane. This is accomplished by means of the Higgs field that transforms as a bi-doublet under $SU(2)_L\times SU(2)_R$. The surviving $SU(2)_V$ implements the custodial symmetry and therefore protects the $T$ parameter \cite{Agashe:2003zs,Csaki:2003zu}. The additional discrete $P_{LR}$ symmetry refers to the exchange of the two $SU(2)$ groups and is important to prevent the left-handed $Zb\bar b$ coupling \cite{Agashe:2006at} from receiving too large corrections. On the UV brane, the symmetry breaking $SU(2)_R\times U(1)_X\to U(1)_Y$ generates the SM gauge group, which is achieved by an interplay between UV and IR boundary conditions. Many technical details of this model can be found in \cite{Casagrande:2010si,Albrecht:2009xr}. For the following analysis we adopt the notations of the first reference.

We start with the relevant Feynman rules needed for the discussion of the decays $h\to VV^*$ in Section~\ref{sec:hVV}. Instead of \eqref{eqn:FRhWW} in the minimal model, the Feynman rules for the $W_\mu^{+(0)} W_\nu^{-(n)}h$ and $Z_\mu^{(0)} Z_\nu^{(n)}h$ vertices read \cite{Hahn:2013nza}
\begin{equation}
\begin{aligned}
   \text{$W$ boson:} & & \frac{2i\tilde m_W^2}{c_{\vartheta_W}^2 v}\,\eta_{\mu\nu}\,
    2\pi\,\vec\chi_0^{\,W}(1)^T \bm{D}_{\vartheta_W}\,\vec\chi_n^{\,W}(1) \,, \\
   \text{$Z$ boson:} & & \frac{i\tilde m_W^2}{c_{\vartheta_W}^2 v}\,\eta_{\mu\nu}\,
    2\pi\,\vec\chi_0^{\,Z}(1)^T \bm{D}_{\vartheta_Z}\,\vec\chi_n^{\,Z}(1) \,,
\end{aligned}
\end{equation}
where we have introduced the matrices (for $V=W,Z$)
\begin{equation}\label{eqn:DW}
   \bm{D}_{\vartheta_V} 
   = \begin{pmatrix} c_{\vartheta_V}^2 & -s_{\vartheta_V} c_{\vartheta_V} \\
    -s_{\vartheta_V} c_{\vartheta_V} & ~s_{\vartheta_V}^2 \end{pmatrix} ,
\end{equation}
with $c_{\vartheta_W}\equiv\cos\vartheta_W=g_{L,5}/\sqrt{g_{L,5}^2+g_{R,5}^2}$ and $s_{\vartheta_W}\equiv\sin\vartheta_W=g_{R,5}/\sqrt{g_{L,5}^2+g_{R,5}^2}$. The 5D gauge couplings $g_{L,5}$ and $g_{R,5}$ belong to the left- and right-handed $SU(2)$ groups. Note that demanding the $P_{LR}$ symmetry fixes $\cos\vartheta_W=\sin\vartheta_W=1/\sqrt2$. The angle $\vartheta_Z$ depends on the 5D gauge couplings in a more complicated way, but under the assumption of the $P_{LR}$ symmetry one finds $\tan^2\vartheta_Z=1-2s_w^2$, where $s_w=\sin\theta_w$ denotes the sine of the Weinberg angle \cite{Casagrande:2010si}. As in the minimal RS model, the parameter $\tilde m_W$ is the leading contribution to the $W$-boson mass in an expansion in powers of $v^2/\mkk^2$. Due to the custodial symmetry in the bulk, this parameter appears in the Higgs coupling to both $W$ and $Z$ bosons. The two-component vectors $\vec \chi_n^W(t)$ and $\vec \chi_n^Z(t)$ contain $Z_2$-even profile functions on the orbifold, whose the upper (lower) components are ``untwisted'' (``twisted'') functions. Untwisted $Z_2$-even functions obey Neumann boundary conditions on the UV brane, allowing for light zero modes. Twisted $Z_2$-even functions obey Dirichlet boundary conditions on the UV brane and are thus not smooth at this orbifold fixed point. Explicitly, the  zero-mode profiles read \cite{Casagrande:2010si}
\begin{equation}\label{eqn:chiWcust}
\begin{split}
\sqrt{2\pi}\,\vec{\chi}_0^{\,W}(t) &= 
\begin{pmatrix} 
1-\frac{m_W^2}{2\mkk^2}\left[t^2 \left(L - \frac{1}{2} +\ln t \right) - \frac{1}{2} + \frac{1}{2L}\right] \\
\frac{L s_{\vartheta_W}}{2 c_{\vartheta_W}} \frac{m_W^2}{\mkk^2} \,t^2
\end{pmatrix} + \ord\left(\frac{v^4}{\mkk^4}\right), \\
\sqrt{2\pi}\,\vec{\chi}_0^{\,Z}(t) & = 
\begin{pmatrix} 
1-\frac{m_Z^2}{2\mkk^2}\left[t^2 \left(L - \frac{1}{2} +\ln t \right) - \frac{1}{2} + \frac{1}{2L}\right] \\
\frac{L s_{\vartheta_Z} c_{\vartheta_Z}}{2 c^2_{\vartheta_W}} \frac{m_W^2}{\mkk^2} \,t^2
\end{pmatrix} + \ord\left(\frac{v^4}{\mkk^4}\right).
\end{split}
\end{equation}
Note that the twisted component is proportional to $t^2$ and suppressed by the ratio $m_W^2/\mkk^2$. It follows that the corrections factors in \eqref{cWres} and \eqref{cZres} become
\begin{equation}
\begin{split}\label{eqn:cWcZ}
   c_W\big|_{\rm cust} 
   &= \frac{v_\SM}{v}\,\frac{\tilde m_W^2}{m_W^2 c_{\vartheta_W}^2}\, 
    2\pi\,\vec\chi_0^{\,W}(1)^T \bm{D}_{\vartheta_W}\,\vec\chi_0^{\,W}(1) 
    = 1 - \frac{m_W^2}{2\mkk^2} \left( 3L - 1 + \frac{1}{2L} \right) +\dots \,, \\
   c_Z\big|_{\rm cust} 
   &= \frac{v_\SM}{v}\,\frac{\tilde m_W^2}{m_Z^2 c_{\vartheta_W}^2}\, 
    2\pi\,\vec\chi_0^{Z}(1)^T \bm{D}_{\vartheta_Z}\,\vec\chi_0^Z(1) \hspace{0.35cm} 
    = 1 - \frac{m_W^2}{2\mkk^2} \left( 3L + 1 - \frac{1}{2L} \right) +\dots \,,
\end{split}
\end{equation}
in accordance with (\ref{eqn:cWZcust}).

The Feynman rules for the couplings of the $W$ and $Z$ bosons and their KK excitations to SM quarks, the $W_\mu^{+(n)}\bar u_A^{(i)} d_A^{(j)}$ and the $Z_\mu^{(n)}\bar q_A^{(i)} q_A^{(i)}$ vertices (with $A=L,R$), are given by 
\begin{align}
\begin{split}
\label{eqn:FRWqqCust}
\text{$W$ boson:} \quad &  \frac{i}{\sqrt 2} \frac{g_{L,5}}{\sqrt{2\pi r} } \int_\e^1 dt \, 
\sqrt{2\pi} \,\, 
\mathcal{U}^{\da(i)}_A (t)
\begin{pmatrix}
 \Omega_W   & \hspace{2mm}
\frac{g_{R,5}}{g_{L,5}}\, \Omega_2    
\end{pmatrix} 
\vec \chi_n^{W}(t)\, 
\ga^\mu \, \mathcal{D}^{(j)}_A(t) P_A\,,\\
\text{$Z$ boson:} \quad & \frac{i}{\sqrt 2} \frac{g_{L,5}}{\sqrt{2\pi r} c_w} \int_\e^1 dt \, 
\sqrt{2\pi} \, 
\mathcal{Q}^{\da(i)}_A (t)
\begin{pmatrix}
 Q_Z^q   & \hspace{2mm}
\frac{g_{Z'\!,5}}{g_{Z,5}}\, Q_{Z'}^q    
\end{pmatrix} 
\vec \chi_n^{Z}(t)\, 
\ga^\mu \, \mathcal{Q}^{(i)}_A(t)P_A\,,
\end{split}
\end{align}
with the chiral projectors $P_{R,L}=\frac{1}{2}(1\pm\ga_5)$. Following \cite{Casagrande:2010si}, we collect all left- and right-handed quark fields in the up, down, and exotic sectors into the 15-component vectors $(\vec U_A,\vec u_A)^T$ and the 9-component vectors $(\vec D_A,\vec d_A)^T$ and $(\vec\Lambda_A,\vec\lambda_A)^T$. We collectively refer to them as ${\cal Q}_{L,R}$, with ${\cal Q} = {\cal U, D}, \Lambda$, defined by 
\begin{equation}\label{QLR}
   \Q_{L,R}(t,x) = \sum_n \Q_{L,R}^{(n)} (t) \, q_{L,R}^{(n)}(x) \,.
\end{equation}
Here $\Q_{L,R}^{(n)}(t)$ are the quark profiles, and $q^{(n)}(x)$ denote the left- and right-handed components of the $n^{\rm th}$ fermion in the KK decomposition. In \eqref{eqn:FRWqqCust} the object $\mathcal{U}^{(n)}_A$ includes the profiles for the $n^\tx{th}$ mode of the five up-type quark fields $(u,u',u^c,U',U)$, where the first two components transform under $SU(2)_L$, while the last three components are $SU(2)_L$ singlets. Likewise $\mathcal{D}^{(n)}_A$ contains the profiles of the down-type quark fields $(d,D',D)$, where only the first field is charged under $SU(2)_L$. The $\Omega_W$ and $\Omega_2$ matrices appearing in \eqref{eqn:FRWqqCust} are $5\times 3$ matrices and given by
\begin{align}
\Omega_W = 
\begin{pmatrix}
 1 & \quad 0 \quad & 0 \\[-1mm] 0 & 0 & 0 \\[-1mm] 0 & 0 & 0 \\[-1mm]
 0 & 0 & 1 \\[-1mm] 0 & 0 & 0 
\end{pmatrix} ,&& 
\Omega_2 = 
\begin{pmatrix}
 0 & \quad 0 \quad & 0 \\[-1mm] 1 & 0 & 0 \\[-1mm] 0 & 0 & 0 \\[-1mm]
 0 & 0 & 0 \\[-1mm] 0 & 1 & 0 
\end{pmatrix}.
\end{align}
Note that for the $W$-boson the leading contribution to the CKM matrix arises from the $(11)$-component of $\Omega_W$. For vertices involving the light SM fermions, corrections coming from the $t$-dependent term in the gauge-boson profile as well as from the admixture of the $U'$ and $D'$ states are chirally suppressed and can be neglected \cite{Casagrande:2010si}. This feature extends to the case of the KK excitations of the $W$ boson. Effectively this means that we only need to keep the constant contributions of the $W$ profiles, which survive near the UV brane and are given by $\vec\chi_n^W(\e)$. In case of the $Z$-boson vertices in the second Feynman rule in \eqref{eqn:FRWqqCust}, we have defined the couplings
\begin{equation}
   \frac{g_{Z'\!,5}^2}{g_{Z,5}^2} 
    = \frac{\cos^2\theta_w \tan^4\vartheta_W}{\tan^2\vartheta_W-\tan^2\theta_w} \,, \qquad
   Q_Z^q = T_L^{q3} - s_w^2 Q_q \,, \qquad
   Q_{Z'} = - T_R^{q3} - \frac{\tan^2\theta_w}{\tan^2\vartheta_W}\,Y^q \,, 
\end{equation}
where $T_{L,R}^{q3}$ denote the eigenvalues under the third generator of $SU(2)_{L,R}$, $Y^q$ is the hyper-charge, and $Q_q$ denotes the electromagnetic charge of the quark. Once again we only need to keep the $t$-independent contributions in the gauge-boson profile functions. Thus, as in the minimal RS model we can approximate the Feynman rules in \eqref{eqn:FRWqqCust} by
\begin{equation}
\begin{aligned}
\text{$W$ boson}: \quad & \frac{i}{\sqrt 2} \frac{g_{5,L}}{\sqrt{2\pi r}} \,\sqrt{2\pi}\,  
\begin{pmatrix}
1 & 0 
\end{pmatrix}\, 
\vec \chi_n^W(\e)\, 
V_{ij}^{\rm CKM} \ga^\mu P_L\,, \\
\text{$Z$ boson}: \quad &  \frac{i}{\sqrt 2} \frac{g_{5,L}}{\sqrt{2\pi r}c_w} \,\sqrt{2\pi}\,  
\begin{pmatrix}
1 & 0 
\end{pmatrix}\, 
\vec \chi_n^Z(\e)\,  \ga^\mu \left[ g_{q,L}(s_w^2) \, P_L + g_{q,R}(s_w^2) \, P_R \right] .
\end{aligned}
\end{equation}
For the SM $W$ and $Z$ bosons ($n=0$), the Feynman rules coincide with the corresponding rules \eqref{eqn:Wqq} and \eqref{eqn:Zff} found in the minimal RS model, since the first components of \eqref{eqn:chiWcust} are the same as the profiles in \eqref{eqn:chiW}. 

Combining all pieces, we find that instead of (\ref{replace2}) we must perform the following replacement in the SM amplitude (with $V=W,Z$):
\begin{equation}\label{eqn:Wudcust}
   \frac{1}{m_V^2-s}
   \to \frac{v_\SM}{v}\,\frac{\tilde m_W^2}{m_V^2 c_{\vartheta_W}^2}\, 
    \sqrt{2\pi}\,\chi_0^{V}(1)^T\,\frac{g_{L,5}}{\sqrt{2\pi r} g}\, 
    2\pi\,\bm{B}^{\rm UV}_V(1,\epsilon;-s)
    \left( \begin{matrix} 1 \\ 0 \end{matrix} \right) .
\end{equation}
The 5D propagator function is defined in terms of the infinite sum
\begin{equation}
   \bm{B}_V^{\rm UV}(t,t';-p^2) = \sum_{n\ge 0}\,
    \frac{\vec{\chi}_n^{\,V}(t)\,\vec{\chi}_n^{\,V}(t')^T}{\left(m^V_n\right)^2-p^2} \,.
\end{equation}
It has been calculated to all orders in $v^2/\mkk^2$ in \cite{Hahn:2013nza}. Expanding the result to first non-trivial order, we obtain 
\begin{align}\label{eqn:Bexpand}
2\pi \,\bs B_V^\UV(t,t';-p^2) =  
\begin{pmatrix}
 \frac{c^V_1(t,t')}{m_V^2-p^2} + \frac{c_2(t,t')}{2\mkk^2}  & \hspace{5mm}\frac{L m_V^2 \tan\vartheta_V}{2 \mkk^2 (m_V^2-p^2)} \, t'^2\vspace{1mm} \\
\frac{L m_V^2 \tan\vartheta_V}{2 \mkk^2 (m_V^2-p^2)} \, t^2 & \frac{L t_<^2}{2 \mkk^2}
\end{pmatrix}  + \ord\left( \frac{v^2}{\mkk^4} \right),
\end{align}
which is valid for momenta $|p^2|\ll\mkk^2$. Here $c_1^V(t,t')=2\pi\,\chi_0^V(t) \chi_0^V(t')$ is defined via the zero-mode profiles of the vector bosons in the minimal RS model, and $c_2(t,t')$ coincides with the expression given in \eqref{eqn:c1c2}. The (11)-component of the propagator is thus the same as in the minimal model. Inserting \eqref{eqn:Bexpand} into \eqref{eqn:Wudcust}, we arrive at \eqref{replace} with $c_W$ and $c_Z$ given by \eqref{eqn:cWcZ}, while 
\begin{equation}
\begin{split}
c_{\Gamma_W}^{1/2}\big|_{\rm cust} &\equiv \frac{g_{L,5}}{\sqrt{2\pi r} g}\, \sqrt{2\pi} \,\left(\begin{matrix} 1 &0 \end{matrix}\right) \vec{ \chi}_0^{\,W}(\epsilon) = 1-\frac{m_W^2}{2\mkk^2}\frac{1}{4L} +\dots \,,\\
c_{\Gamma_Z}^{1/2}\big|_{\rm cust}& \equiv  \frac{g_{L,5}}{\sqrt{2\pi r} g}\, \sqrt{2\pi} \, \left(\begin{matrix} 1 &0 \end{matrix}\right) \vec{\chi}_0^Z(\epsilon) \hspace{0.15cm}= c_{\Gamma_W}^{1/2} \left[1+\frac{m_Z^2-m_W^2}{4\mkk^2}\left(1-\frac{1}{L} \right)+\dots  \right]  
\end{split}
\end{equation}
remain the same as in the minimal model, see (\ref{666}) and (\ref{777}).

The vector-boson fusion process analyzed in Section~\ref{sec:VBF} can be studied analogously. In this case, we need to replace the first line of \eqref{eqn:AWWh} by
\begin{align}
   & \frac{1}{(m_V^2-p_1^2)\,(m_V^2-p_2^2)} \\
   &\qquad \to \frac{v_{\rm SM}}{v}\,\frac{\tilde m_W^2}{m_V^2 c_{\vartheta_W}^2}
    \left( \frac{g_{L,5}}{\sqrt{2\pi r} g} \right)^2 (2\pi)^2
    \left( \begin{matrix} 1 & 0 \end{matrix}\right) \bs B^\UV_V(\e,1;-p_1^2)\, 
    \bs D_{\vartheta_V}\,\bs B^\UV_V(1,\e;-p_2^2) \begin{pmatrix} 1 \\ 0 \end{pmatrix} .
    \notag 
\end{align}
Using the expansions for the propagator functions and evaluating the rescaling factors, we confirm the second line of \eqref{eqn:AWWh} with $c_V$ and $c_{\Gamma_V}^{1/2}$ given above.

\end{appendix}

%\newpage


\begin{thebibliography}{99}

\bibitem{ATLAS:2012gk} 
  G.~Aad {\it et al.}  [ATLAS Collaboration],
  %``Observation of a new particle in the search for the Standard Model Higgs boson with the ATLAS detector at the LHC,''
  Phys.\ Lett.\ B {\bf 716}, 1 (2012)
  [arXiv:1207.7214 [hep-ex]].
  %%CITATION = ARXIV:1207.7214;%%
  
\bibitem{CMS:2012gu} 
  S.~Chatrchyan {\it et al.}  [CMS Collaboration],
  %``Observation of a new boson at a mass of 125 GeV with the CMS experiment at the LHC,''
  Phys.\ Lett.\ B {\bf 716}, 30 (2012)
  [arXiv:1207.7235 [hep-ex]].
  %%CITATION = ARXIV:1207.7235;%%  

\bibitem{Randall:1999ee}  
  L.~Randall and R.~Sundrum,  
  %``A large mass hierarchy from a small extra dimension,''  
  Phys.\ Rev.\ Lett.\  {\bf 83}, 3370 (1999)  
  [hep-ph/9905221].  
  %%CITATION = HEP-PH 9905221;%%

\bibitem{Grossman:1999ra}
  Y.~Grossman and M.~Neubert,
  %``Neutrino masses and mixings in non-factorizable geometry,''
  Phys.\ Lett.\ B {\bf 474}, 361 (2000)
  [hep-ph/9912408].
  %%CITATION = HEP-PH 9912408;%%

\bibitem{Gherghetta:2000qt}
  T.~Gherghetta and A.~Pomarol,
  %``Bulk fields and supersymmetry in a slice of AdS,''
  Nucl.\ Phys.\ B {\bf 586}, 141 (2000)
  [hep-ph/0003129].
  %%CITATION = HEP-PH 0003129;%%

\bibitem{Huber:2000ie}
  S.~J.~Huber and Q.~Shafi,
  %``Fermion masses, mixings and proton decay in a Randall-Sundrum model,''
  Phys.\ Lett.\ B {\bf 498}, 256 (2001)
  [hep-ph/0010195].
  %%CITATION = HEP-PH 0010195;%%

\bibitem{Agashe:2004ay} 
  K.~Agashe, G.~Perez and A.~Soni,
  %``B-factory signals for a warped extra dimension,''
  Phys.\ Rev.\ Lett.\  {\bf 93}, 201804 (2004)
  [hep-ph/0406101].
  %%CITATION = HEP-PH/0406101;%%

\bibitem{Agashe:2004cp} 
  K.~Agashe, G.~Perez and A.~Soni,
  %``Flavor structure of warped extra dimension models,''
  Phys.\ Rev.\ D {\bf 71}, 016002 (2005)
  [hep-ph/0408134].
  %%CITATION = HEP-PH/0408134;%%
  
\bibitem{Csaki:2008zd}
  C.~Csaki, A.~Falkowski and A.~Weiler,
  %``The Flavor of the Composite Pseudo-Goldstone Higgs,''
  JHEP {\bf 0809}, 008 (2008)
  [arXiv:0804.1954 [hep-ph]].
  %%CITATION = JHEPA,0809,008;%%  

\bibitem{Casagrande:2008hr}
  S.~Casagrande, F.~Goertz, U.~Haisch, M.~Neubert and T.~Pfoh,
  %``Flavor Physics in the Randall-Sundrum Model: I. Theoretical Setup and
  %Electroweak Precision Tests,''
  JHEP {\bf 0810}, 094 (2008)
  [arXiv:0807.4937 [hep-ph]].
  %%CITATION = JHEPA,0810,094;%%  

\bibitem{Blanke:2008zb} 
  M.~Blanke, A.~J.~Buras, B.~Duling, S.~Gori and A.~Weiler,
  %``$\Delta$ F=2 Observables and Fine-Tuning in a Warped Extra Dimension with Custodial Protection,''
  JHEP {\bf 0903}, 001 (2009)
  [arXiv:0809.1073 [hep-ph]].
  %%CITATION = ARXIV:0809.1073;%%

\bibitem{Blanke:2008yr} 
  M.~Blanke, A.~J.~Buras, B.~Duling, K.~Gemmler and S.~Gori,
  %``Rare K and B Decays in a Warped Extra Dimension with Custodial Protection,''
  JHEP {\bf 0903}, 108 (2009)
  [arXiv:0812.3803 [hep-ph]].
  %%CITATION = ARXIV:0812.3803;%%

\bibitem{Bauer:2009cf} 
  M.~Bauer, S.~Casagrande, U.~Haisch and M.~Neubert,
  %``Flavor Physics in the Randall-Sundrum Model: II. Tree-Level Weak-Interaction Processes,''
  JHEP {\bf 1009}, 017 (2010)
  [arXiv:0912.1625 [hep-ph]].
  %%CITATION = ARXIV:0912.1625;%%

\bibitem{Heinemeyer:2013tqa} 
  S.~Heinemeyer {\it et al.}  [LHC Higgs Cross Section Working Group Collaboration],
  %``Handbook of LHC Higgs Cross Sections: 3. Higgs Properties,''
  arXiv:1307.1347 [hep-ph].
  %%CITATION = ARXIV:1307.1347;%%

\bibitem{Djouadi:2007fm} 
  A.~Djouadi and G.~Moreau,
  %``Higgs production at the LHC in warped extra-dimensional models,''
  Phys.\ Lett.\ B {\bf 660}, 67 (2008)
  [arXiv:0707.3800 [hep-ph]].
  %%CITATION = ARXIV:0707.3800;%%

\bibitem{Falkowski:2007hz}
  A.~Falkowski,
  %``Pseudo-Goldstone Higgs Production via Gluon Fusion,''
  Phys.\ Rev.\  D {\bf 77}, 055018 (2008)
  [arXiv:0711.0828 [hep-ph]].
  %%CITATION = PHRVA,D77,055018;%%

\bibitem{Cacciapaglia:2009ky} 
  G.~Cacciapaglia, A.~Deandrea and J.~Llodra-Perez,
  %``Higgs ---> Gamma Gamma beyond the Standard Model,''
  JHEP {\bf 0906}, 054 (2009)
  [arXiv:0901.0927 [hep-ph]].
  %%CITATION = ARXIV:0901.0927;%%}

\bibitem{Bhattacharyya:2009nb} 
  G.~Bhattacharyya and T.~S.~Ray,
  %``Probing warped extra dimension via gg ---> h and h ---> gamma gamma at LHC,''
  Phys.\ Lett.\ B {\bf 675}, 222 (2009)
  [arXiv:0902.1893 [hep-ph]].
  %%CITATION = ARXIV:0902.1893;%%

\bibitem{Casagrande:2010si}
  S.~Casagrande, F.~Goertz, U.~Haisch, M.~Neubert and T.~Pfoh,
  %``The Custodial Randall-Sundrum Model: From Precision Tests to Higgs
  %Physics,''
  JHEP {\bf 1009}, 014 (2010)
  [arXiv:1005.4315 [hep-ph]].
  %%CITATION = JHEPA,1009,014;%%
  
\bibitem{Goertz:2011hj} 
  F.~Goertz, U.~Haisch and M.~Neubert,
  %``Bounds on Warped Extra Dimensions from a Standard Model-like Higgs Boson,''
  Phys.\ Lett.\ B {\bf 713}, 23 (2012)
  [arXiv:1112.5099 [hep-ph]].
  %%CITATION = ARXIV:1112.5099;%%

\bibitem{Azatov:2010pf}
  A.~Azatov, M.~Toharia and L.~Zhu,
  %``Higgs Production from Gluon Fusion in Warped Extra Dimensions,''
  Phys.\ Rev.\  D {\bf 82}, 056004 (2010)
  [arXiv:1006.5939 [hep-ph]].
  %%CITATION = PHRVA,D82,056004;%%
  
\bibitem{Malm:2013jia} 
  R.~Malm, M.~Neubert, K.~Novotny and C.~Schmell,
  %``5D Perspective on Higgs Production at the Boundary of a Warped Extra Dimension,''
  JHEP {\bf 1401}, 173 (2014)
  [arXiv:1303.5702 [hep-ph]].
  %%CITATION = ARXIV:1303.5702,;%%

\bibitem{Carena:2012fk} 
  M.~Carena, S.~Casagrande, F.~Goertz, U.~Haisch and M.~Neubert,
  %``Higgs Production in a Warped Extra Dimension,''
  JHEP {\bf 1208}, 156 (2012)
  [arXiv:1204.0008 [hep-ph]].
  %%CITATION = ARXIV:1204.0008;%%
  
\bibitem{Bouchart:2009vq} 
  C.~Bouchart and G.~Moreau,
  %``Higgs boson phenomenology and VEV shift in the RS scenario,''
  Phys.\ Rev.\ D {\bf 80}, 095022 (2009)
  [arXiv:0909.4812 [hep-ph]].
  %%CITATION = ARXIV:0909.4812;%%
  
  
\bibitem{Peskin:2012we} 
  M.~E.~Peskin,
  %``Comparison of LHC and ILC Capabilities for Higgs Boson Coupling Measurements,''
  arXiv:1207.2516 [hep-ph].
  %%CITATION = ARXIV:1207.2516;%%

\bibitem{Klute:2013cx} 
  M.~Klute, R.~Lafaye, T.~Plehn, M.~Rauch and D.~Zerwas,
  %``Measuring Higgs Couplings at a Linear Collider,''
  Europhys.\ Lett.\  {\bf 101}, 51001 (2013)
  [arXiv:1301.1322 [hep-ph]].
  %%CITATION = ARXIV:1301.1322;%%
 
\bibitem{Asner:2013psa} 
  D.~M.~Asner, T.~Barklow, C.~Calancha, K.~Fujii, N.~Graf, H.~E.~Haber, A.~Ishikawa and S.~Kanemura {\it et al.},
  %``ILC Higgs White Paper,''
  arXiv:1310.0763 [hep-ph].
  %%CITATION = ARXIV:1310.0763;%%

\bibitem{Tian:2013yda} 
  J.~Tian {\it et al.}  [ILD Collaboration],
  %``Measurement of Higgs couplings and self-coupling at the ILC,''
  PoS EPS-HEP2013, 316 (2013)
  [arXiv:1311.6528 [hep-ph]].
  %%CITATION = ARXIV:1311.6528;%%
    
\bibitem{Baer:2013cma} 
  H.~Baer, T.~Barklow, K.~Fujii, Y.~Gao, A.~Hoang, S.~Kanemura, J.~List and H.~E.~Logan {\it et al.},
  %``The International Linear Collider Technical Design Report - Volume 2: Physics,''
  arXiv:1306.6352 [hep-ph].
  %%CITATION = ARXIV:1306.6352;%%

\bibitem{Agashe:2003zs}
  K.~Agashe, A.~Delgado, M.~J.~May and R.~Sundrum,
  %``RS1, custodial isospin and precision tests,''
  JHEP {\bf 0308}, 050 (2003)
  [hep-ph/0308036].
  %%CITATION = HEP-PH 0308036;%%

\bibitem{Csaki:2003zu} 
  C.~Csaki, C.~Grojean, L.~Pilo and J.~Terning,
  %``Towards a realistic model of Higgsless electroweak symmetry breaking,''
  Phys.\ Rev.\ Lett.\  {\bf 92}, 101802 (2004)
  [hep-ph/0308038].
  %%CITATION = HEP-PH/0308038;%%

\bibitem{Agashe:2006at}
  K.~Agashe, R.~Contino, L.~Da Rold and A.~Pomarol,
  %``A custodial symmetry for Z b anti-b,''
  Phys.\ Lett.\ B {\bf 641}, 62 (2006)
  [hep-ph/0605341].
  %%CITATION = HEP-PH 0605341;%%

\bibitem{Hahn:2013nza} 
  J.~Hahn, C.~H\"orner, R.~Malm, M.~Neubert, K.~Novotny and C.~Schmell,
  %``Higgs Decay into Two Photons at the Boundary of a Warped Extra Dimension,''
  Eur.\ Phys.\ J.\ C {\bf 74}, 2857 (2014)
  [arXiv:1312.5731 [hep-ph]].
  %%CITATION = ARXIV:1312.5731;%%
  
\bibitem{Archer:2014jca} 
  P.~R.~Archer, M.~Carena, A.~Carmona and M.~Neubert,
  %``Higgs Production and Decay in Models of a Warped Extra Dimension with a Bulk Higgs,''
  [arXiv:1408.5406 [hep-ph]].
  %%CITATION = ARXIV:1408.5406;%%

\bibitem{BulkHiggs} 
  C.~H\"orner, R.~Malm, M.~Neubert and C.~Schmell, 
  in preparation.

\bibitem{Keung:1984hn} 
  W.~-Y.~Keung and W.~J.~Marciano,
  %``HIGGS SCALAR DECAYS: H ---> W+- X,''
  Phys.\ Rev.\ D {\bf 30}, 248 (1984).
  %%CITATION = PHRVA,D30,248;%%

\bibitem{Randall:2001gb} 
  L.~Randall and M.~D.~Schwartz,
  %``Quantum field theory and unification in AdS5,''
  JHEP {\bf 0111}, 003 (2001)
  [hep-th/0108114].
  %%CITATION = HEP-TH/0108114;%%
  
\bibitem{Csaki:2002gy} 
  C.~Csaki, J.~Erlich and J.~Terning,
  %``The Effective Lagrangian in the Randall-Sundrum model and electroweak physics,''
  Phys.\ Rev.\ D {\bf 66}, 064021 (2002)
  [hep-ph/0203034].
  %%CITATION = HEP-PH/0203034;%%

\bibitem{Carena:2003fx} 
  M.~S.~Carena, A.~Delgado, E.~Ponton, T.~M.~P.~Tait and C.~E.~M.~Wagner,
  %``Precision electroweak data and unification of couplings in warped extra dimensions,''
  Phys.\ Rev.\ D {\bf 68}, 035010 (2003)
  [hep-ph/0305188].
  %%CITATION = HEP-PH/0305188;%%
  
  \bibitem{Fichet:2013ola} 
  S.~Fichet and G.~von Gersdorff,
  %``Anomalous gauge couplings from composite Higgs and warped extra dimensions,''
  JHEP {\bf 1403}, 102 (2014)
  [arXiv:1311.6815 [hep-ph]].
  %%CITATION = ARXIV:1311.6815;%%
  %6 citations counted in INSPIRE as of 28 Aug 2014

\bibitem{Beneke:2002jn} 
  M.~Beneke and M.~Neubert,
  %``Flavor singlet B decay amplitudes in QCD factorization,''
  Nucl.\ Phys.\ B {\bf 651}, 225 (2003)
  [hep-ph/0210085].
  %%CITATION = HEP-PH/0210085;%%

\bibitem{Djouadi:2005gj} 
  A.~Djouadi,
  %``The Anatomy of electro-weak symmetry breaking. II. The Higgs bosons in the minimal supersymmetric model,''
  Phys.\ Rept.\  {\bf 459}, 1 (2008)
  [hep-ph/0503173].
  %%CITATION = HEP-PH/0503173;%%
  
\bibitem{Albrecht:2009xr} 
  M.~E.~Albrecht, M.~Blanke, A.~J.~Buras, B.~Duling and K.~Gemmler,
  %``Electroweak and Flavour Structure of a Warped Extra Dimension with Custodial Protection,''
  JHEP {\bf 0909}, 064 (2009)
  [arXiv:0903.2415 [hep-ph]].
  %%CITATION = ARXIV:0903.2415;%%

\bibitem{Carena:2006bn} 
  M.~S.~Carena, E.~Ponton, J.~Santiago and C.~E.~M.~Wagner,
  %``Light Kaluza Klein States in Randall-Sundrum Models with Custodial SU(2),''
  Nucl.\ Phys.\ B {\bf 759}, 202 (2006)
  [hep-ph/0607106].
  %%CITATION = HEP-PH/0607106;%%
    
\bibitem{Cacciapaglia:2006gp} 
  G.~Cacciapaglia, C.~Csaki, G.~Marandella and J.~Terning,
  %``A New custodian for a realistic Higgsless model,''
  Phys.\ Rev.\ D {\bf 75}, 015003 (2007)
  [hep-ph/0607146].
  %%CITATION = HEP-PH/0607146;%%

\bibitem{Contino:2006qr} 
  R.~Contino, L.~Da Rold and A.~Pomarol,
  %``Light custodians in natural composite Higgs models,''
  Phys.\ Rev.\ D {\bf 75}, 055014 (2007)
  [hep-ph/0612048].
  %%CITATION = HEP-PH/0612048;%%

\bibitem{Davoudiasl:1999tf} 
  H.~Davoudiasl, J.~L.~Hewett and T.~G.~Rizzo,
  %``Bulk gauge fields in the Randall-Sundrum model,''
  Phys.\ Lett.\ B {\bf 473}, 43 (2000)
  [hep-ph/9911262].
  %%CITATION = HEP-PH/9911262;%%

\bibitem{Brod:2013cka} 
  J.~Brod, U.~Haisch and J.~Zupan,
  %``Constraints on CP-violating Higgs couplings to the third generation,''
  JHEP {\bf 1311}, 180 (2013)
  [arXiv:1310.1385 [hep-ph]].
  %%CITATION = ARXIV:1310.1385;%%

\bibitem{Baron:2013eja} 
  J.~Baron {\it et al.}  [ACME Collaboration],
  %``Order of Magnitude Smaller Limit on the Electric Dipole Moment of the Electron,''
  Science {\bf 343}, no. 6168, 269 (2014)
  [arXiv:1310.7534 [physics.atom-ph]].
  %%CITATION = ARXIV:1310.7534;%%
  
\bibitem{Bishara:2013vya} 
  F.~Bishara, Y.~Grossman, R.~Harnik, D.~J.~Robinson, J.~Shu and J.~Zupan,
  %``Probing CP Violation in $h\rightarrow\gamma\gamma$ with Converted Photons,''
  JHEP {\bf 1404}, 084 (2014)
  [arXiv:1312.2955 [hep-ph]].
  %%CITATION = ARXIV:1312.2955;%%
  
\bibitem{ATLASMoriond2014}
  ATLAS Collaboration,
  ATLAS-CONF-2014-009.

\bibitem{CMSMoriond2013}
  CMS Collaboration,
  CMS-PAS-HIG-13-005.

%\bibitem{Barger:2012hv} 
%  V.~Barger, M.~Ishida and W.~-Y.~Keung,
 % %``Total Width of 125 GeV Higgs Boson,''
 % Phys.\ Rev.\ Lett.\  {\bf 108}, 261801 (2012)
 % [arXiv:1203.3456 [hep-ph]].
  %%CITATION = ARXIV:1203.3456;%%

\bibitem{Denner:2011mq} 
  A.~Denner, S.~Heinemeyer, I.~Puljak, D.~Rebuzzi and M.~Spira,
  %``Standard Model Higgs-Boson Branching Ratios with Uncertainties,''
  Eur.\ Phys.\ J.\ C {\bf 71}, 1753 (2011)
  [arXiv:1107.5909 [hep-ph]].
  %%CITATION = ARXIV:1107.5909;%%
  %119 citations counted in INSPIRE as of 20 Aug 2014

\bibitem{ATLAS-CONF-2013-079}
  ATLAS Collaboration,
  ATLAS-CONF-2013-079.
 
\bibitem{Chatrchyan:2013zna} 
  S.~Chatrchyan {\it et al.}  [CMS Collaboration],
  %``Search for the standard model Higgs boson produced in association with a W or a Z boson and decaying to bottom quarks,''
  Phys.\ Rev.\ D {\bf 89}, 012003 (2014)
  [arXiv:1310.3687 [hep-ex]].
  %%CITATION = ARXIV:1310.3687;%%

\bibitem{Chatrchyan:2014nva} 
  S.~Chatrchyan {\it et al.}  [CMS Collaboration],
  %``Evidence for the 125 GeV Higgs boson decaying to a pair of $\tau$ leptons,''
  JHEP {\bf 1405}, 104 (2014)
  [arXiv:1401.5041 [hep-ex]].
  %%CITATION = ARXIV:1401.5041;%%

\bibitem{Blanke:2012tv} 
  M.~Blanke, B.~Shakya, P.~Tanedo and Y.~Tsai,
  %``The Birds and the Bs in RS: The $b to s \gamma$ penguin in a warped extra dimension,''
  JHEP {\bf 1208}, 038 (2012)
  [arXiv:1203.6650 [hep-ph]].
  %%CITATION = ARXIV:1203.6650;%%
  %18 citations counted in INSPIRE as of 04 Jul 2014

\bibitem{Contino:2003ve} 
  R.~Contino, Y.~Nomura and A.~Pomarol,
  %``Higgs as a holographic pseudoGoldstone boson,''
  Nucl.\ Phys.\ B {\bf 671}, 148 (2003)
  [hep-ph/0306259].
  %%CITATION = HEP-PH/0306259;%%

\bibitem{Agashe:2004rs} 
  K.~Agashe, R.~Contino and A.~Pomarol,
  %``The Minimal composite Higgs model,''
  Nucl.\ Phys.\ B {\bf 719}, 165 (2005)
  [hep-ph/0412089].
  %%CITATION = HEP-PH/0412089;%%

\end{thebibliography}
\end{document}